\documentclass{aa}

\usepackage{psfig}
\usepackage{graphicx}
\usepackage{natbib}
\usepackage[english]{babel}  
\usepackage{txfonts}

\begin{document}

\title{Shock-cloud interaction in the Vela SNR II.\\Hydrodynamic model}

\author{M. Miceli\inst{1,2} \and F. Reale\inst{1,2} \and S. Orlando\inst{2} \and F. Bocchino\inst{2}}

\offprints{M. Miceli,\\ \email{miceli@astropa.unipa.it}}

\institute{Dipartimento di Scienze Fisiche ed Astronomiche, Sezione di Astronomia, Universit\`a di Palermo, Piazza del Parlamento 1, 90134 Palermo, Italy
\and 
INAF - Osservatorio Astronomico di Palermo, Piazza del Parlamento 1, 90134 Palermo, Italy
}

\date{Received, accepted}

\authorrunning{M. Miceli et al.}
\titlerunning{Modelling the Vela FilD}

\abstract
{In the framework of the study of the X-ray and optical emission in supernova remnants we focus on an isolated X-ray knot in the northern rim of the Vela SNR (Vela FilD), whose X-ray emission has been studied and discussed in Paper I.}
{We aim at understanding the physical origin of the X-ray and optical emission in FilD, at understanding the role of the different physical processes at work, and at obtaining a key for the interpretation of future X-ray observations of SNRs.}
{To this end we have pursued an accurate ``forward'' modeling of the interaction of the Vela SNR shock with an ISM cloud. We perform hydrodynamic simulations and we directly compare the observables synthesized from the simulations with the data.}
{We explore four different model setups, choosing the values of the physical parameters on the basis of our preliminary analysis of the X-ray data. We synthesize X-ray emission maps and spectra filtered through the \emph{XMM-Newton} EPIC-MOS instrumental response. The impact of a shock front at 6 million Kelvin on an elliptical cloud 30 times denser than the ambient medium describes well the shock-cloud interaction processes in the Vela FilD region in terms of spectral properties and morphology of the X-ray and optical emission.}
{The bulk of the X-ray emission in the FilD knot originates in the cloud material heated by the transmitted shock front, but significant X-ray emission is also associated to the cloud material, which evaporates, as an effect of thermal conduction, in the intercloud medium. The physical origin of the FilD optical emission is associated to thermal instabilities. In the FilD knot the X-ray emission associated to the reflected shock front is negligible.}

\keywords{Hydrodynamics -- Shock waves --  ISM: supernova remnants -- ISM: clouds -- ISM: kinematics and dynamics -- ISM: individual object: Vela SNR}

\maketitle

\section{Introduction}
\label{Introduction}

The study of supernova remnants (SNRs) allows us to probe the physical and chemical properties of the interstellar medium (ISM). Moreover, SNRs drive the mass and energy exchange in the ISM, thus playing a fundamental role for its evolution.

The X-ray emission of SNRs is characterized by a clumpy morphology that has been interpreted as a result of the interaction between the blast wave expanding shock and inhomogeneities of the ambient medium. The impact of the shock front with ISM clouds influences the direction and the speed of the SNR expansion and determines the heating and the ionization of the clouds. Several authors (see, for example, \citealt{dsb97}, \citealt{bms00}, and \citealt{pfr02}) have shown how the study of X-ray and optical observations of SNRs provides important information about the shock-cloud interaction process and about the physical and chemical conditions of the post-shock medium (temperature, chemical composition, morphology of the clouds, etc.). However, the details of the heating processes and the physical scenario which determines the observed emission are still not well understood.

In this framework, shock-cloud interaction has been extensively studied both analytically (e. g. \citealt{mc75}, \citealt{hb83}) and numerically (e. g. \citealt{bw90}, \citealt{sn92}, \citealt{kmc94}, \citealt{xs95}, and \citealt{fag05}) and, recently, the combined effects of thermal conduction and radiative cooling have been investigated for the first time by \citet{opr05}. This work has shown that, when radiative losses dominates, the shocked cloud is fragmented into cold and dense cloudlets, while thermal conduction determines the evaporation of the clouds and suppresses hydrodynamic instabilities. Despite these large theoretical efforts, however, no detailed comparison with the observations has yet been provided. In fact, although comparisons with the observations are present, for example, in  \cite{hs84} and in \cite{wl91}, these models are very simplified and do not include at the same time all the relevant physical effects. Recently, an observation-oriented numerical model has been presented by \citet{pf05} to describe the shock-cloud interaction in the southwestern region of the Cygnus Loop. However, no observable quantities are produced from this model and only a simple qualitative comparison between the model density plot and the overall morphology of the H$\alpha$ emission is presented.

\citet{mbm05} (hereafter Paper I) analyzed an \emph{XMM-Newton} EPIC observation of a small isolated knot (named FilD) in the northern rim of the Vela SNR. 
The analysis revealed a link between optical and X-ray emission and provided information about density, temperature, extension along the line of sight, and chemical composition of the FilD cloud. In particular, the spectral analysis has shown that the X-ray emission of FilD is described well by an optically-thin plasma with two thermal components at $\sim 1\times 10^{6}$ K and $\sim 3\times 10^{6}$ K, respectively. The inhomogeneity of the surface brightness in the FilD region is determined by different volume distributions along the line of sight of the two components.

Here we present a detailed hydrodynamic model of the shock-cloud interaction in the Vela FilD. Our approach consists in synthesizing observable quantities from the model (for example, maps of X-ray surface brightness and spectra), comparing them directly with the \emph{XMM-Newton} EPIC data and with optical observations of the FilD region. We also perform on the synthesized products the same analysis we carried on for the EPIC data.  The aims of this work are: i) the proper interpretation of the X-ray observation (i. e. to understand the physical origin of the two spectral components, to ascertain the evolutionary stage of the interaction, to obtain information about the density and temperature profiles of the emitting plasma); ii) to understand the role of the different physical processes (transmitted and reflected shocks, thermal conduction, radiative cooling) in the evolution of the system; iii) to obtain a key for the interpretation of future X-ray observations of middle-aged SNRs.
 
The paper is organized as follows: in Sect. \ref{FilD} we review the main results of Paper I that give us constraints for our simulations; the model equations and setups are presented in Sect. \ref{The Model} and in Sect. \ref{The model setup}; our results are presented in Sect. \ref{Results} and discussed in Sect. \ref{Discussion}. Finally, our conclusions are drawn in Sect. \ref{Summary and conclusions}.

\section{Hydrodynamic modeling}

\subsection{Constraints from the observations}
\label{FilD}

\begin{figure*}[htb!]
 \centerline{\hbox{
     \psfig{figure=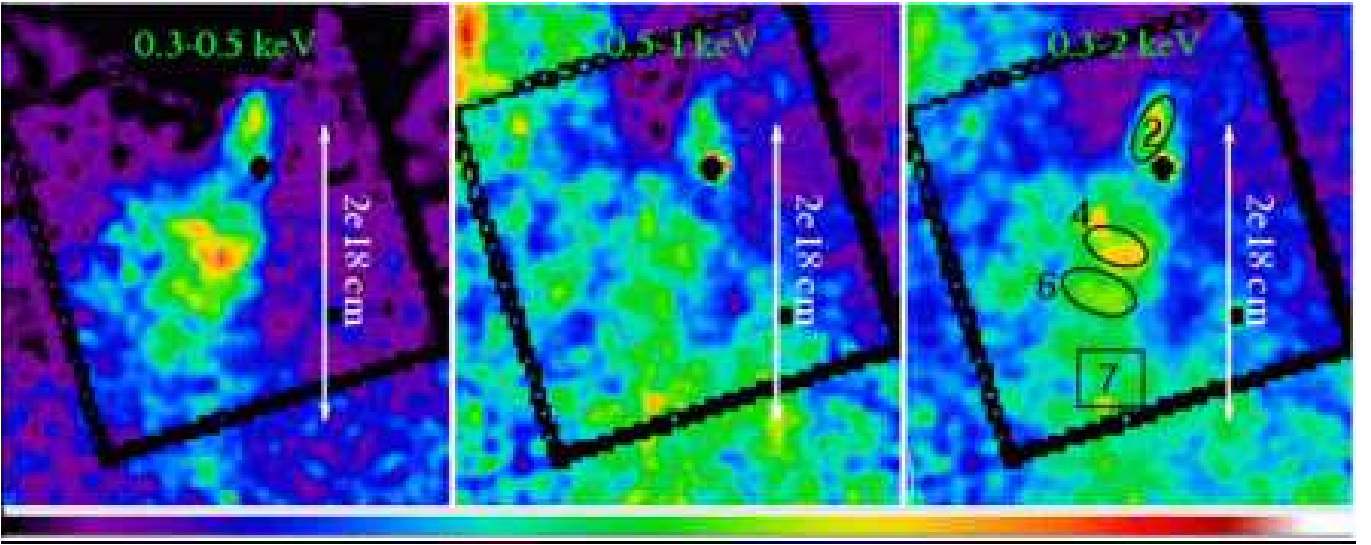,width=\textwidth}	 
     }}
 \centerline{}  	
 \caption{\emph{XMM-Newton} EPIC MOS count rate images of the Vela FilD knot, observed in different energy bands. We smoothed the images adaptively to a signal-to-noise ratio of 10. The bin size is $10.4''$, corresponding to $3.9\times 10^{16}$ cm, assuming a distance $D=250$ pc. The color bar has a linear scale and the count rates range between $1.2\times10^{-4}$ s$^{-1}$ bin$^{-1}$ and $6\times 10^{-4}$ s$^{-1}$ bin$^{-1}$, in the $0.3-0.5$ keV and $0.5-1$ keV images, and between $2.4\times 10^{-4}$ s$^{-1}$ bin$^{-1}$ and $1.2\times10^{-3}$ s$^{-1}$ bin$^{-1}$, in the $0.3-2$ keV images.}
 \label{fig:XMM}
 \end{figure*}

Fig. \ref{fig:XMM} shows the \emph{XMM-Newton} EPIC MOS count rate images of the FilD region in the $0.3-0.5$ keV, $0.5-1$ keV, and $0.3-2$ keV energy bands. The bin size is $10.4''$ and the images are adaptively smoothed to a signal-to-noise ratio of 10 and vignetting-corrected as in \citet{mdb06}. The \emph{XMM-Newton} images show that shape and size of the FilD X-ray knot are similar in the three energy bands.
In Paper I we found that the extension of the X-ray emitting knot is $\sim (2\times 1\times 1)\times 10^{18}$ cm. 

\begin{figure}[htb!]
 \centerline{\hbox{
 \psfig{figure=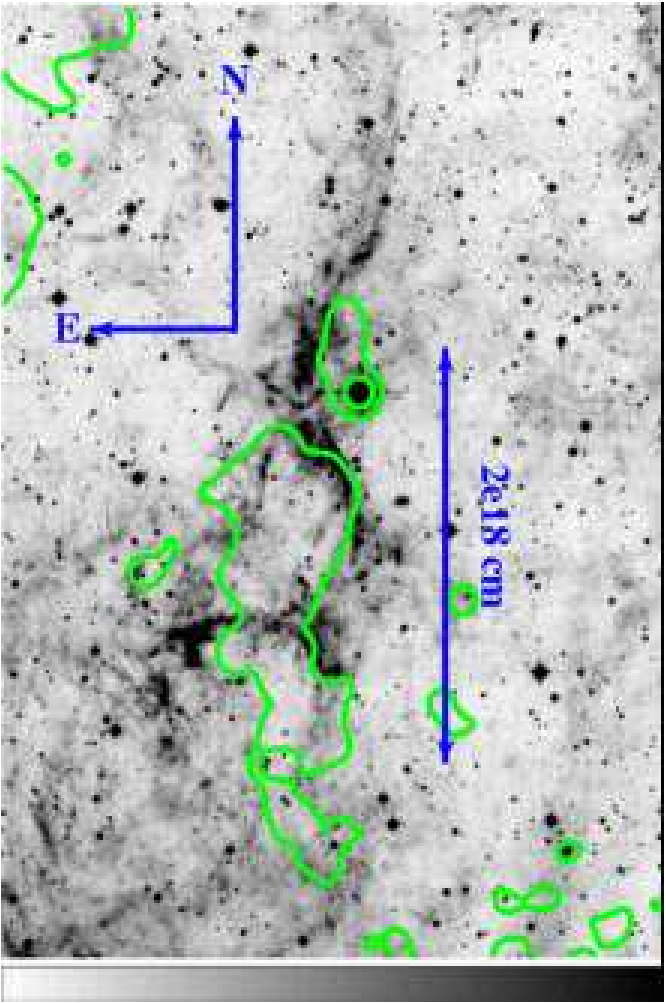,width=\columnwidth}
       }}
\caption{Vela FilD H$\alpha$ emission (from AAO/UKST H-alpha survey). We have superimposed, in green, the X-ray contour levels at a count rate 2.4 times higher than the minimum, derived from the EPIC observation in the $0.3-2$ keV band (Paper I, Fig. 2, left panel).}
\label{fig:X-Halpha}
\end{figure}

The Vela FilD emits both in the X-ray and optical bands. Fig. \ref{fig:X-Halpha} shows an H$\alpha$ image of the FilD region, including the X-ray contour levels. A bright optical filament spans between the FilD regions with the highest X-ray surface brightness. The filament is almost aligned along the South-North direction and, therefore, it is perpendicular to the plane of the incident shock front. Optical filaments in SNRs are usually associated to the plasma behind the transmitted shock which travels through dense interstellar clouds, because they are typically aligned with the transmitted shock front planes; the peculiar orientation of the FilD optical filament is difficult to explain according to this scenario.

As discussed in Paper I, the spectral analysis of the \emph{XMM-Newton} EPIC data shows that the X-ray emission in the FilD region is described well by two thermal components in collisional ionization equilibrium (CIE) at $\sim 1\times 10^{6}$ K and $\sim 3\times 10^{6}$ K, respectively. The emission measure ($EM$) of the cooler component is about an order of magnitude larger than that of the hotter component (Paper I, Fig.8). We associated these components with two different phases of the clouds: the cloud core (which corresponds to the cooler component), surrounded by a hotter and less dense corona. The core post-shock density is $n_{core}=1.5-5$ cm$^{-3}$ and the corona post-shock density is about three times lower. Moreover, we calculated the distance of the FilD cloud from the center of the Vela shell and, by using the Sedov model (\citealt{sed59}), we estimated that the shock velocity in the intercloud medium is $5.7-7.8\times10^{7}$ cm$/$s. Therefore, following \citet{mh80}, the shock temperature is $T_{sh}=4.6-8.6\times10^{6}$ K. 
We also derived the mean abundances of O ($\overline{O}/O_\odot\approx 1.0$), Fe ($\overline{Fe}/Fe_\odot =0.39\pm 0.05$), and Ne ($\overline{Ne}/Ne_\odot=1.7\pm 0.2$). We found that  the inhomogeneity of the surface brightness in the FilD region is determined by different volume distributions along the line of sight of the two components. In particular, the surface brightness increases with the $EM$ per unit area of the soft component. Finally, we estimated that FilD was shocked $\sim 1300-4500$ yr ago.
We will use these results as constraints for our modeling.

\subsection{The model equations}
\label{The Model}

\begin{figure}[htb!]
 \centerline{\hbox{
     \psfig{figure=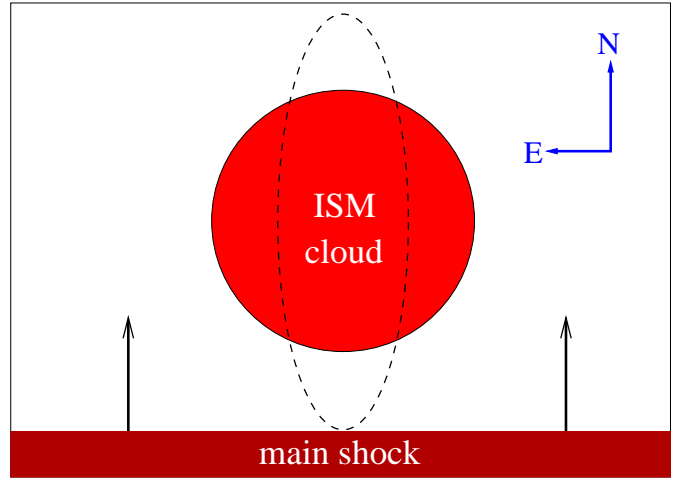,width=\columnwidth}	 
     }}
 \caption{Schematic representation of the initial conditions of our simulations. The system consists of a planar shock-front which interacts with a spherical$/$elliptical (dashed line) isolated ISM cloud.}
 \label{fig:initial}
 \end{figure}

We model the impact of a plane-parallel SNR shock on an isolated uniform cloud in pressure equilibrium with the ambient (intercloud) isothermal medium (Fig.\ref{fig:initial}). The plane-parallel approach is justified because the FilD cloud is small ($<1$ pc) with respect to the curvature radius of the Vela shell ($\sim15$ pc). The shock-cloud interaction is described by means of a two-dimensional hydrodynamic model. The model solves the time-dependent compressible fluid equations of mass, momentum, and energy conservation, including radiative losses from an optically thin thermal plasma and thermal conduction (considering also the effects of heat flux saturation). The model equations are:
\begin{equation}
\frac{\partial \rho}{\partial t} + \nabla \cdot \rho \mbox{\bf v} = 0
\label{eq:massa}
\end{equation}

\begin{equation}
\frac{\partial \rho \mbox{\bf v}}{\partial t} +\nabla \cdot \rho
\mbox{\bf vv} + \nabla P = 0
\label{eq:momento}
\end{equation}

\begin{equation}
\frac{\partial \rho E}{\partial t} +\nabla\cdot (\rho E+P)\mbox{\bf v}
= -\nabla\cdot \textbf{q} -n_e n_H \Lambda(T)
\label{eq:energia}
\end{equation}

\noindent with

\begin{equation}
E = \epsilon +\frac{1}{2} |\mbox{\bf v}|^2~,~~~~~P=(\gamma-1)\rho\epsilon~,
\end{equation}

\noindent where {\bf v} is the bulk velocity of the gas, $t$ is the time, $P$ the pressure, $T$ the temperature, $n_e$ and $n_H$ are the electron and hydrogen number density, respectively (we assume $n_{e}=n_{H}$), $\rho=\mu m_{H}n_{H}$ is the mass density ($\mu=1.26$, assuming cosmic abundances), $\Lambda(T)$ is the radiative losses function per unit emission measure  (\citealt{rs77}, \citealt{mgv85}), and $q$ is the conductive flux, defined following \citet{db93} as $q=(1/q_{spi}+1/q_{sat})^{-1}$, where $q_{spi}$ and $q_{sat}$ are the classical (\citealt{spi62}) and saturated (\citealt{db93}) conductive flux, respectively
\begin{equation}
q_{spi}=-\kappa (T)\nabla T~{\rm ,} ~~~~~~~
q_{sat}={\rm -sign}(\nabla T)5\phi \rho c_{s}^{3}
\end{equation}
with $\kappa(T)=5.6\times10^{-7}T^{5/2}$ erg s$^{-1}$ K$^{-1}$ cm$^{-1}$, $\phi=0.3$ (in agreement with the values suggested by \citealt{giu84} and \citealt{bsm89} for a fully ionized cosmic gas), and where $c_{s}$ is the isothermal sound speed.

We solved the set of equations by using the FLASH code (\citealt{for00}), upgraded to include thermal conduction and radiative losses from an optically thin thermal plasma (see \citealt{opr05}). The FLASH code is based on a directionally split PPM Riemann solver (\citealt{wc84}), and uses adaptive mesh refinement (PARAMESH, \citealt{mom00}), particularly appropriate to describe the moving steep gradients at the boundaries of the shock front. 
Our simulations have been performed on an IBM pSeries p4 machine and on a Linux cluster of AMD opteron processors, for a total of $\sim 9000$ CPU hours.

\subsection{The model setup}
\label{The model setup}

\begin{center}
\begin{table*}[htb!]
\begin{center}
\caption{Physical parameters of the model setups. In all the setups the intercloud pre-shock temperature is $10^{4}$ K and the intercloud pre-shock density is $0.05$ cm$^{-3}$. The setups marked in boldface are those discussed in detail in this paper. }
\begin{tabular}{lccccc} 
\hline\hline
Model setup & Cloud morphology  &  Cloud Radius$/$semiaxis & Cloud density     & Cloud mass         &  Shock temperature \\ 
            &                   &      ($10^{18}$ cm)       &     (cm$^{-3}$)  &   (M$_\odot$)      &     ($10^{6}$ K)    \\ \hline
\textbf{Sph1} &\textbf{Sphere} &   {\bf \emph{R}$_{cl}=$~2}     & {\bf \emph{n}$_{cl}=$~1}  & {\bf\emph{M}$_{cl}=$~0.035} & {\bf \emph{T}$_{sh}=$~4.6}  \\ 
Sph2    & Sphere            &      $R_{cl}=3.09$       & $n_{cl}=1.5$      & $M_{cl}=0.196$     & $T_{sh}=6$          \\
Ell1    & Ellipsoid      & $a_{cl}=c_{cl}=1$, $b_{cl}=3.09$ & $n_{cl}=1$   & $M_{cl}=0.032$ & $T_{sh}=4.6$        \\
\textbf{Ell2}& \textbf{Ellipsoid}& {\bf\emph{a}$_{cl}=$~\emph{c}$_{cl}=$~1}, {\bf\emph{b}$_{cl}=$~3.09} & {\bf\emph{n}$_{cl}=$~1.5} & {\bf\emph{M}$_{cl}=$~0.047} & {\bf\emph{T}$_{sh}=$~6}\\
\hline\hline
\label{tab:setups}
\end{tabular}
\end{center}
\end{table*}
\end{center}
We show results of four simulations with different model setups. The setup of the physical parameters is dictated by the results of our analysis of the X-ray data. The relevant parameters are: the cloud$/$intercloud density contrast, the shock temperature, and the cloud dimension.
Table \ref{tab:setups} shows the initial physical parameters of our four setups. In all the setups the intercloud pre-shock temperature is $10^{4}$ K and the intercloud pre-shock density is $0.05$ cm$^{-3}$. In setup Sph1 and Sph2 the cloud is spherical, whereas in setup Ell1 and Ell2 we consider an ellipsoidal cloud, with the semimajor axis $b_{cl}$ aligned with the incident shock velocity. The Mach number $\mathcal{M}$ (i. e. the shock speed in units of the intercloud isothermal sound speed) and the density contrast $\chi$ (i. e. the ratio cloud$/$intercloud density) of our setups are $\mathcal{M}_{S1E1}=50$, $\chi_{S1E1}=20$ (for setup Sph1 and Ell1); $\mathcal{M}_{S2E2}=57$, $\chi_{S2E2}=30$ (for setup Sph2 and setup Ell2). According to \citet{opr05} (Fig. 2), for these values radiative cooling dominates over thermal conduction. This implies the local formation of thermal instabilities in structures with dimensions $L>L_{F}=1.3\times 10^{6}T^{2}/n$ ($L_{F}\sim10^{16}$ cm, for $T=2\times10^{5}$ K and $n=5$ cm$^{-3}$), which is the critical Field length (\citealt{fie65}). Notice, however, that the results reported in \citet{opr05} were derived for an impact of a SNR shock with a spherical cloud with radius $R=3.09\times10^{18}$ cm, while our clouds are smaller. So we expect lower values of the characteristic length of temperature variations $l_{T}$ in our simulations. Since the characteristic conductive time-scale is 
\begin{equation}
\tau_{cond}\sim\frac{7Pl_{T}^{2}}{2(\gamma -1)\kappa(T)T}~\rm{,}
\end{equation}
smaller clouds imply a higher efficiency of thermal conduction, which may inhibit the formation of thermal instabilities.

The simulations are performed in a cylindrical 2-D coordinate system ($r,~ z$), assuming axial symmetry. The computational domain extends over $1.5\times10^{19}$ cm in the $r$ direction and over $2\times10^{19}$ cm in the $z$ direction. We use reflection boundary conditions at $r=0$ and zero-gradient boundary conditions (for $\bf{v}$, $\rho$, and $p$) elsewhere. The shock velocity is parallel to the $z$ axis, which corresponds to the North direction in the observation analyzed in Paper I. The post-shock initial conditions are given in the strong shock limit (\citealt{zr67}). The finest spatial resolution is $\sim1.95\times10^{16}$ cm, therefore we have more than 100 zones per cloud radius in setup Sph1 and Sph2 and $\ga150$ and $\ga 50$ zones per semimajor and semiminor axis, respectively, in setup Ell1 and Ell2.

\subsection{Synthesis of the X-ray emission}
\label{Xsynth}
Our modeling allows us to simulate the detailed evolution of the temperature and density of the shock-cloud system. From the computed temperature and density we are able to synthesize the X-ray emission detectable with the \emph{XMM-Newton} EPIC-MOS1 camera (\citealt{taa01}). By rotating the 2D maps of $n$ and $T$ about the symmetry axis, we obtain the full 3D distributions in the cartesian space ($x',~y',~z'$), where the $y'$ axis corresponds to the direction of the line of sight and is perpendicular to the $(r,~z)$ plane. We then derive the emission measure, $EM(x',~y',~z')$, and temperature, $T(x',~y',~z')$, for each fluid element and the distribution $EM(x',~z')$ vs. $T(x',~z')$ (hereafter $EMD$) integrated along the line of sight, for each $(x',~z')$, in the temperature range $0.58-16\times 10^{6}$ K (using 145 bins equispaced in $\log T$). From the EMDs, we synthesize the maps of the X-ray emission and the focal plane spectra using the MEKAL spectral code (\citealt{mgv85}, \citealt{mlv86}, \citealt{log95}), with solar abundances for all the elements, but Fe ($Fe/Fe_\odot =0.39$) and Ne ($Ne/Ne_\odot=1.7$), in agreement with the results of the data analysis (see Sect. \ref{FilD}). The source is assumed at a distance $D=250$ pc, corresponding to the best estimate of the distance of the Vela SNR, which is known with a $\sim30\%$ uncertainty (see Paper I and references therein). We filtered the emission through the \emph{XMM-Newton} EPIC-MOS1 instrumental response and the interstellar column density at the Vela FilD knot, $N_{H}=1\times 10^{20}$ cm$^{-2}$ (see Paper I and references therein). The exposure time ($t_{exp}^{synth}=50$ ks, for the MOS1 camera) is the same as that of the observation analyzed in Paper I ($t_{exp}^{obs}\sim 25$ ks for each MOS camera). 

We have integrated the emission over a cube with edge length $10^{19}$ cm (centered on the cloud). This length corresponds to the maximum extension of the intercloud post-shock medium along the line of sight and was chosen so as to match the minima of the X-ray synthesized surface brightness with those observed with \emph{XMM-Newton}. 

The X-ray emission maps have been produced in the same energy bands analyzed in Paper I (i. e. $0.3-0.5$ keV, $0.5-1$ keV, and $0.3-2$ keV) and are presented in Sect. \ref{A: spherical cloud} and Sect. \ref{B: ellipsoidal cloud}. The space binning of all the model X-ray maps is $3.9\times 10^{16}$ cm that corresponds to $10.4''$ (i. e. the bin-size of the \emph{XMM-Newton} EPIC images of Fig. \ref{fig:XMM}) at 250 pc. 
Focal plane spectra have been synthesized for different regions of the domain, to perform a spatially resolved spectral analysis. All spectra have been analyzed using XSPEC and the chosen on-axis response file is m1$\_$medv9q20t5r6$\_$all$\_$15.rsp.

\section{Results}
\label{Results}

\subsection{The spherical cloud}
\label{A: spherical cloud}

\begin{figure}[htb!]
 \centerline{\hbox{
     \psfig{figure=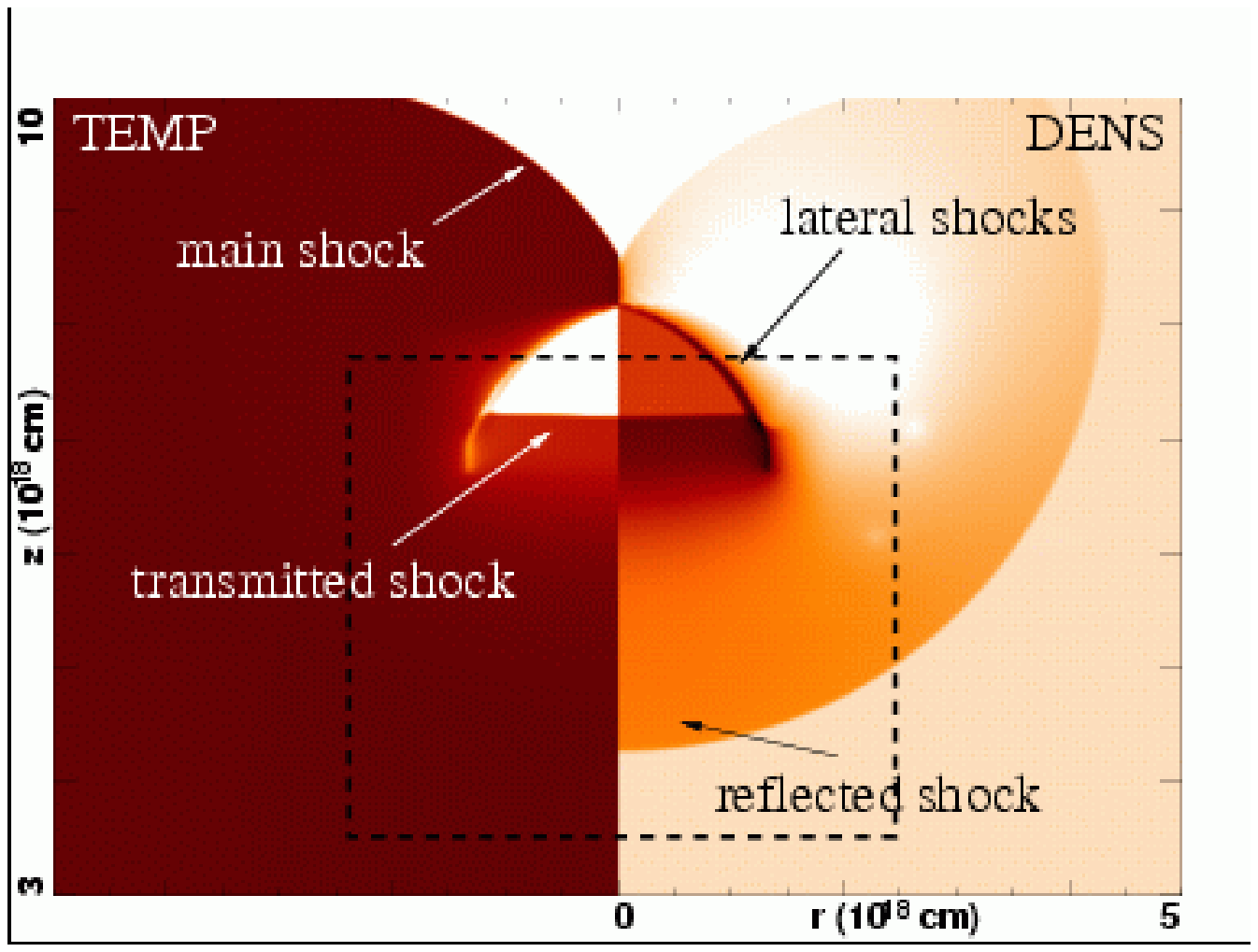,width=\columnwidth}     
     }}
 \centerline{}       
 \centerline{}      
 \centerline{\hbox{          
     \psfig{figure=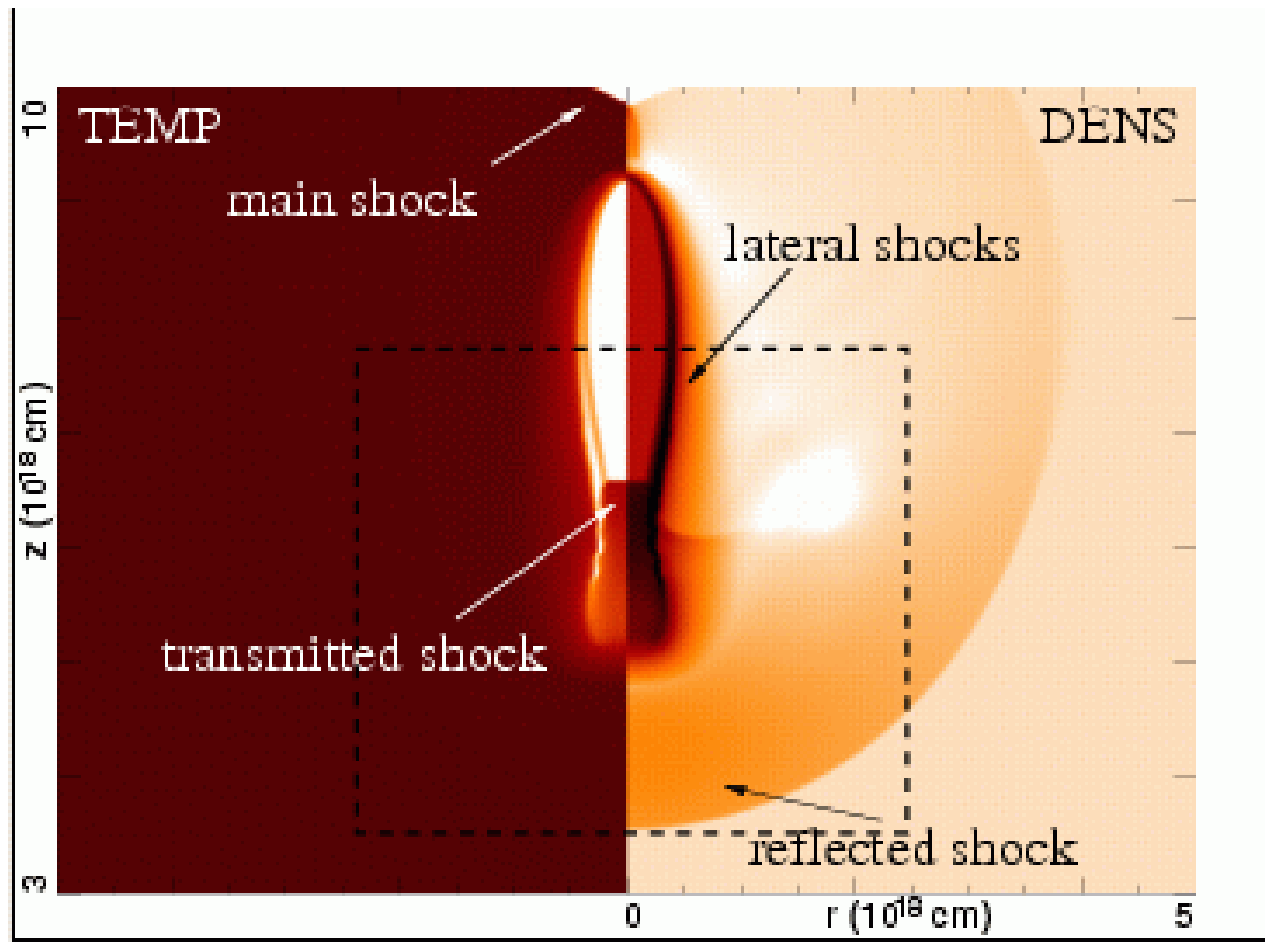,width=\columnwidth}     
     }}    
 \centerline{}      
 \centerline{}    
 \centerline{}       
 \centerline{}      
 \centerline{\hbox{          
    \psfig{figure=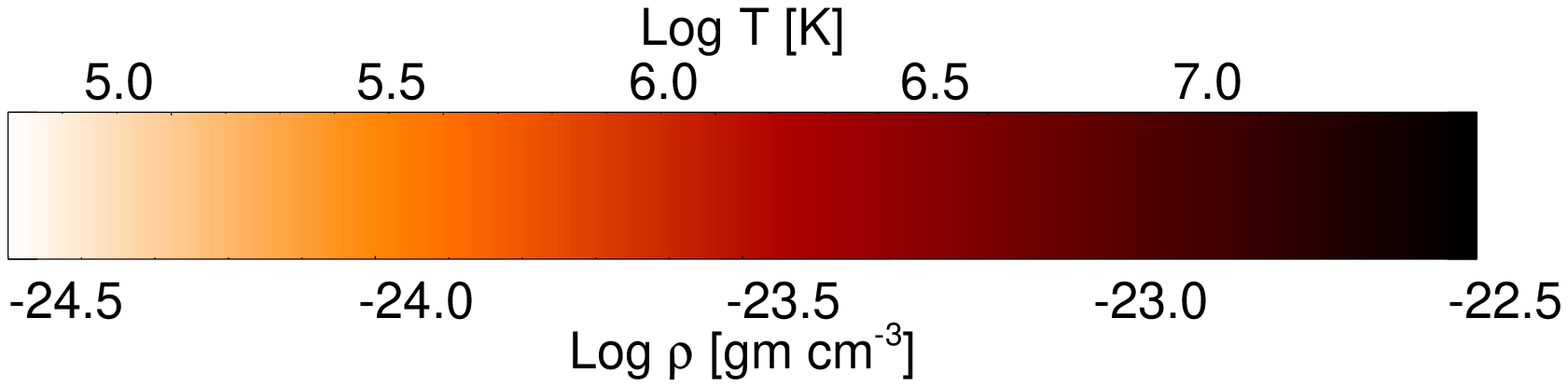,angle=270,width=\columnwidth}     
     }}
 \centerline{}          
 \caption{\emph{Upper panel:} temperature (on the left) and density (on the right) 2D cross-sections through the $(r,~z)$ plane 3400 yr after the impact of the shock front with the spherical cloud (setup Sph1). \emph{Lower panel:} same as upper panel for the ellipsoidal cloud (setup Ell2), 3550 yr after the shock impact. The color bar has a logarithmic scale. The dashed box indicates the field of view of the X-ray count-rate maps showed in Fig. \ref{fig:mappeXAB}.}
 \label{fig:nTAB}
 \end{figure}

The upper panel of Fig. \ref{fig:nTAB} shows, for \textbf{setup Sph1}, the 2D cross-sections through the $(r,~z)$ plane of temperature and density 3400 yr after the first impact of the shock front with the spherical cloud. This stage of the interaction is the one which best approaches the observed data, as shown in Appendix A. The temperature and density cross-sections at different interaction times and the corresponding synthesized X-ray maps are available as online material.

As the shock interacts with the cloud, part of it is transmitted through the cloud (with temperature $\sim 10^{6}$ K and density $\sim 4$ cm$^{-3}$) and a reflected shock propagates backwards in the ambient medium. 
Figure \ref{fig:nTAB} also shows that the plasma behind the reflected bow shock has a density $n\sim 0.6$ cm$^{-3}$ and it is about three times denser than the intercloud post-shock plasma. The pressure behind the transmitted shock front, $P^{t}$, the reflected shock front, $P^{r}$, and the main shock front, $P^{ms}$,  are: $P^{t}\la 1\times10^{-9}$ dyne cm$^{-2}$, $P^{r}\sim 6.5\times10^{-10}$ dyne cm$^{-2}$, and $P^{ms}\sim 2.5\times10^{-10}$ dyne cm$^{-2}$. The thermal conduction between the cloud and the intercloud medium drives a thin hot halo around the cloud. As the primary shock progressively engulfs the cloud, it generates lateral shocks propagating into the cloud.

After $\sim3000$ yr, in the regions where the transmitted shock interacts with the lateral shocks, there are dense ($n\sim 5$ cm$^{-3}$), cold ($T\sim1-2\times 10^{5}$ K), and elongated ($L\sim 5\times 10^{17}$ cm) structures. These structures are larger than the critical Field length and are therefore thermally instable. They cool off to temperatures of a few $10^{4}$ K, reaching density $n\sim 10$ cm$^{-3}$ in a few $10^{2}$ yr (and are visible in Fig. \ref{fig:nTAB}).  
These values are consistent with the characteristic temperature and density of optical filaments in SNRs and with those observed in the FilD region (\citealt{bms00}). We caution the reader that our simulations model a plasma where hydrogen is fully ionized and that for very low values of temperature our radiative losses function is not calculated accurately. Therefore the values of temperature and density derived when $T$ approaches the H ionization temperature may not be completely correct. However, our model clearly indicates that thermal instabilities can produce the FilD optical emission. 
\begin{figure*}[htb!]
 \centerline{\hbox{
     \psfig{figure=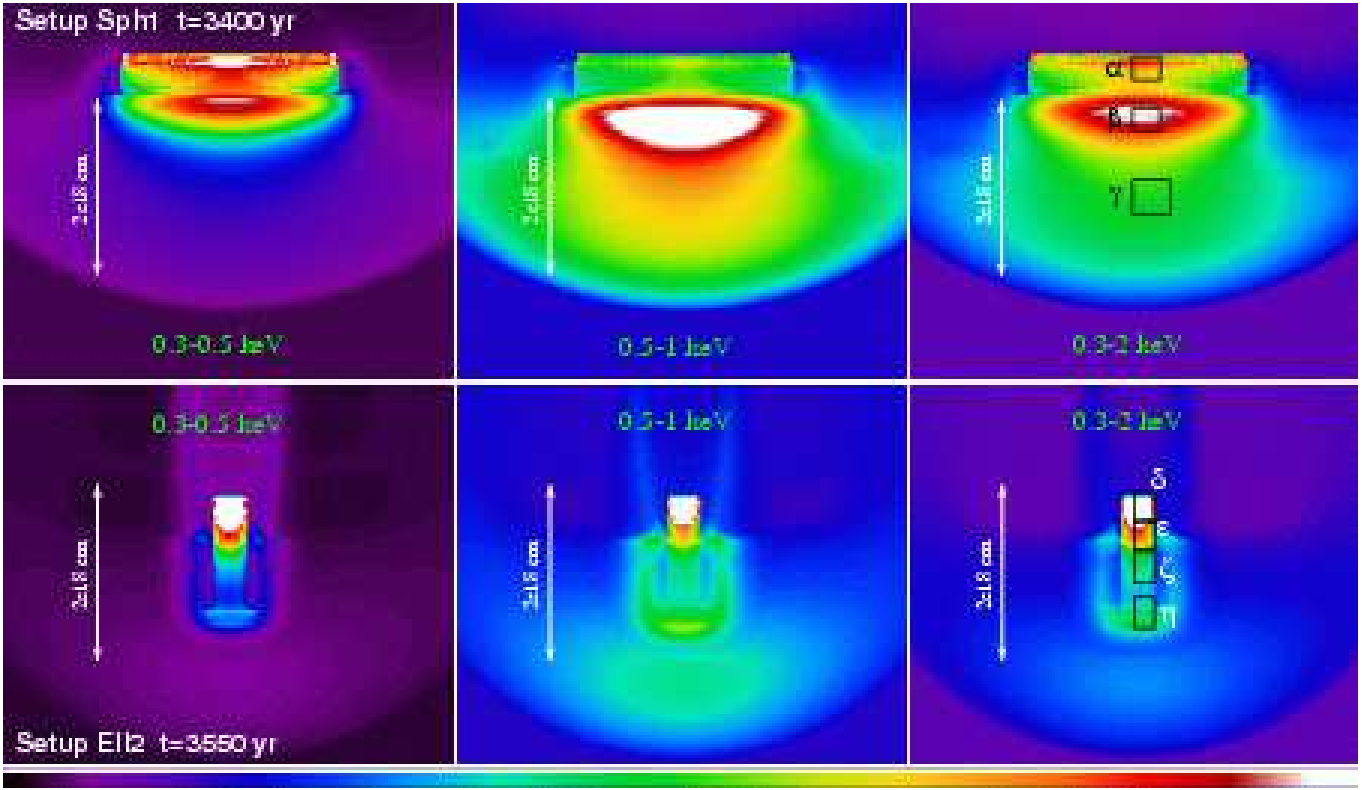,width=\textwidth}     
     }}
 \centerline{}          
 \caption{Synthesized \emph{XMM-Newton} EPIC MOS count rate images in different energy bands for the interaction of the Vela SNR shock front with a spherical cloud at $t=3400$ yr after the first impact of the shock with the cloud (setup Sph1, \emph{upper panels}) and with an ellipsoidal cloud at $t=3550$ yr (setup Ell2, \emph{lower panels}). The field of view of these image is indicated by a dashed box in Fig. \ref{fig:nTAB}. The bin size of the images is $3.9\times 10^{16}$ cm. The color bar has a linear scale between 0 and 
$1.2\times 10^{-3}$ s$^{-1}$ bin$^{-1}$, in the $0.3-0.5$ keV and $0.5-1$ keV images, and between 0 and 
$2.4\times 10^{-3}$ s$^{-1}$ bin$^{-1}$, in the $0.3-2$ keV images. The count rate values and the $2\times10^{18}$ cm scale have been obtained assuming a distance $D=250$ pc. The boxes indicates the regions for the spatially resolved spectral analysis (see Table \ref{tab:spettriAB}).}
 \label{fig:mappeXAB}
 \end{figure*}

\begin{figure}[htb!]
 \centerline{\hbox{
 \psfig{figure=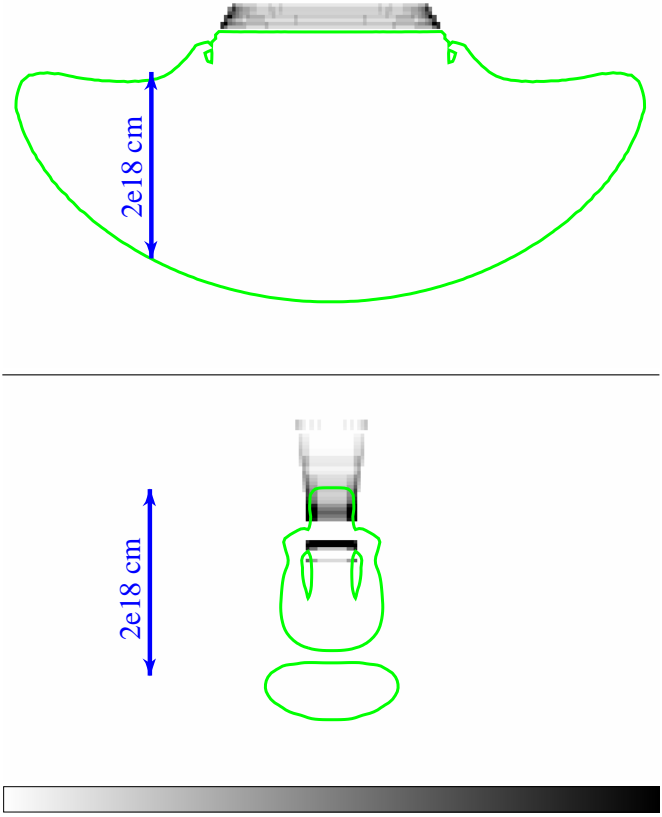,width=\columnwidth}
       }}
\caption{Synthesized map (projected in the plane of the sky) of the emission measure of the plasma with $T<10^{5}$ K and $n>3$ cm$^{-3}$, integrated along the line of sight, 3400 yr after the impact of the Vela shock front with a spherical cloud (setup Sph1, upper panel) and 3550 yr after the impact with an ellipsoidal cloud (setup Ell2, lower panel). As in Fig. \ref{fig:X-Halpha}, we have superimposed, in green, the X-ray contour levels at a count rate 2.4 times higher than the minimum, derived from the synthesized X-ray images in the $0.3-2$ keV band shown in Fig. \ref{fig:mappeXAB}.}
\label{fig:X-HalphaS}
\end{figure}

The upper panels in Fig. \ref{fig:mappeXAB} show the synthesized X-ray maps, for setup Sph1, at the same epoch of Fig. \ref{fig:nTAB}, derived in the three different energy bands, $0.3-0.5$ keV, $0.5-1$ keV, and $0.3-2$ keV, by following the procedure described in Sect. \ref{Xsynth}.  The synthesized count rate per space bin is in agreement with the observed one, considering its strong dependence on the source distance $D$ (the synthesized count rate scales as $D^{-4}$, because the flux scales as $D^{-2}$ and the angular extension of the bin as $D^{-1}\times D^{-1}$) and that the distance of the Vela SNR is known with $\sim30\%$ uncertainty. Although the overall morphology of the synthesized emission in the $0.3-2$ keV band for setup Sph1 appears similar to the observed one, there are significant differences. Great part of the synthesized emission above 0.5 keV, shown in the central upper panel of Fig. \ref{fig:mappeXAB}, is clearly associated to the plasma heated by the reflected shock (see Fig. \ref{fig:nTAB}), while the bulk of the soft X-ray emission in the $0.3-0.5$ keV band originates in the cloud plasma heated by the transmitted shock. The morphology of the synthesized emission below $0.5$ keV is then very different from the synthesized emission in the $0.5-1$ keV band. In setup Sph1, therefore, the morphology of the X-ray emitting knot in the broad band $0.3-2$ keV is the result of the superposition of the emission associated with two different physical conditions of the plasma: the cloud material heated by the transmitted shock at North, which contributes with soft emission, and the intercloud material heated by the reflected shock in the central and southern parts of the knot, with harder emission. The extension of the X-ray emitting knot in the $0.3-2$ keV band is significantly larger than that in the $0.3-0.5$ keV band. This result does not match the \emph{XMM-Newton} observation of the FilD knot, which shows that the X-ray emission in the $0.3-0.5$ keV band has almost the same extension as the ones in the $0.5-1$ keV and $0.3-2$ keV bands (Fig. \ref{fig:XMM}).

The upper panel of Fig. \ref{fig:X-HalphaS} shows, for setup Sph1, the emission measure of the plasma with $T<10^{5}$ K and $n>3$ cm$^{-3}$ projected in the plane of the sky, which is a proxy of the optical emission. We also superimpose, in green, the corresponding X-ray contour levels at a count rate 2.4 times higher than the minimum, derived from the synthesized X-ray images in the $0.3-2$ keV band shown in Fig. \ref{fig:mappeXAB}. The impact of the Vela shock front with the spherical cloud of setup Sph1 produces an optical filament aligned with the incident shock front, at odd with the observation (Fig. \ref{fig:X-Halpha}).
\begin{figure}[htb!]
 \centerline{\hbox{
     \psfig{figure=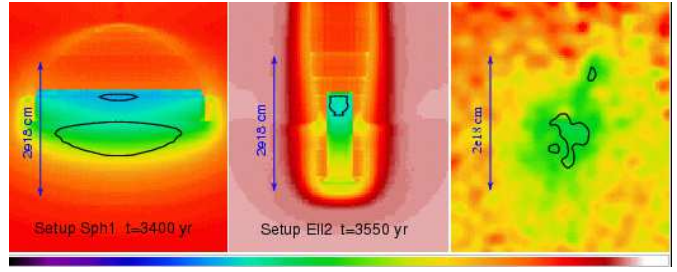,width=\columnwidth}	 
     }}
 \centerline{}  	
 \caption{Synthesized mean photon energy map for setup Sph1 (3400 yr after the shock impact), \emph{left panel}, setup Ell2 (3550 yr after the shock impact), \emph{central panel}, and observed mean photon energy map in the Vela FilD region, \emph{right panel}. Each pixel holds the mean energy of the EPIC MOS photons in the $0.3-2$ keV energy band. The color scale is linear and ranges between 300 eV and 870 eV. We superimposed (in white) the corresponding contour level at $75\%$ of the maximum count rate in the same energy band.}
 \label{fig:avgEAB}
 \end{figure}

\begin{figure}[htb!]
 \centerline{\hbox{
     \psfig{figure=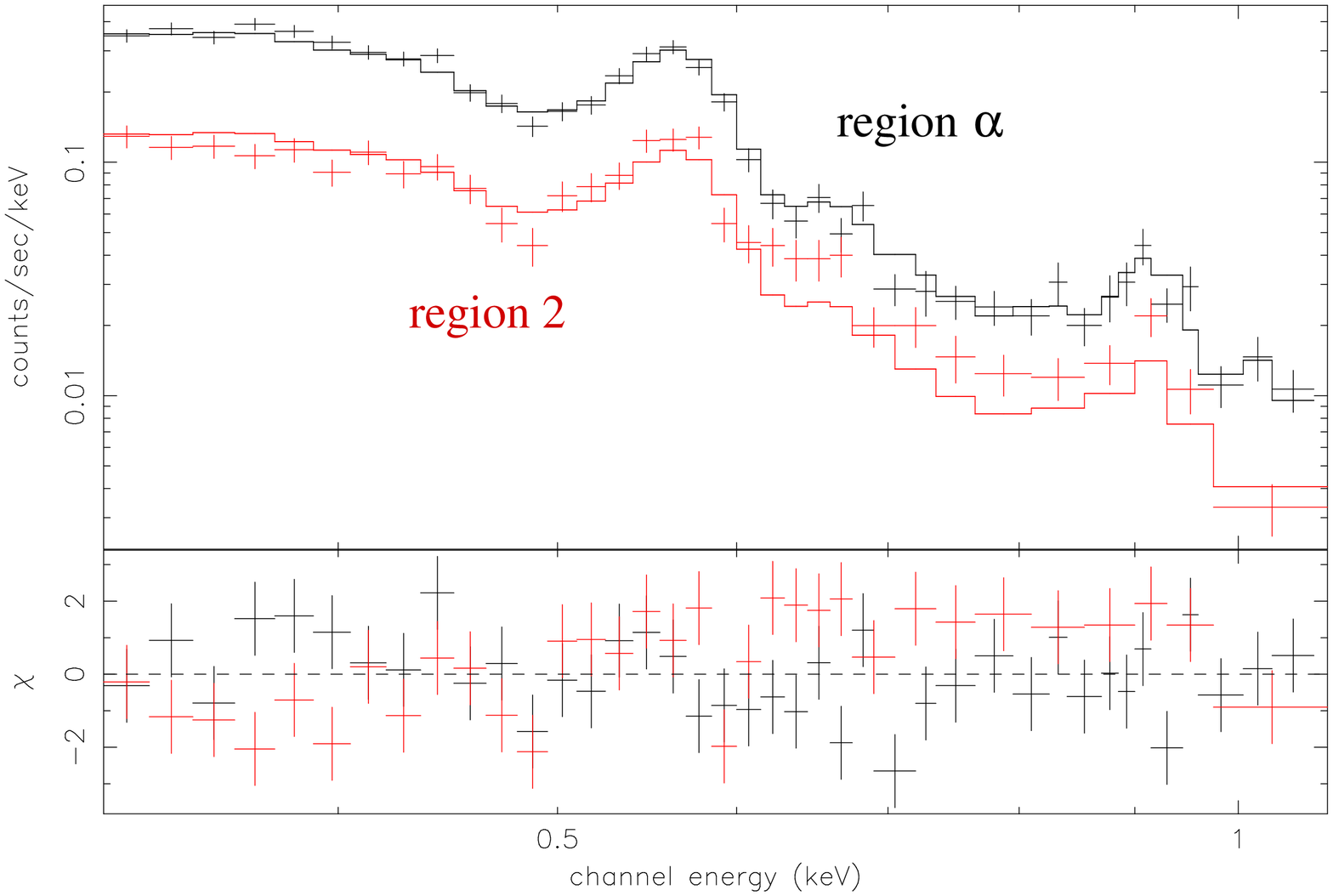,width=\columnwidth}	 
     }}
 \centerline{\hbox{
     \psfig{figure=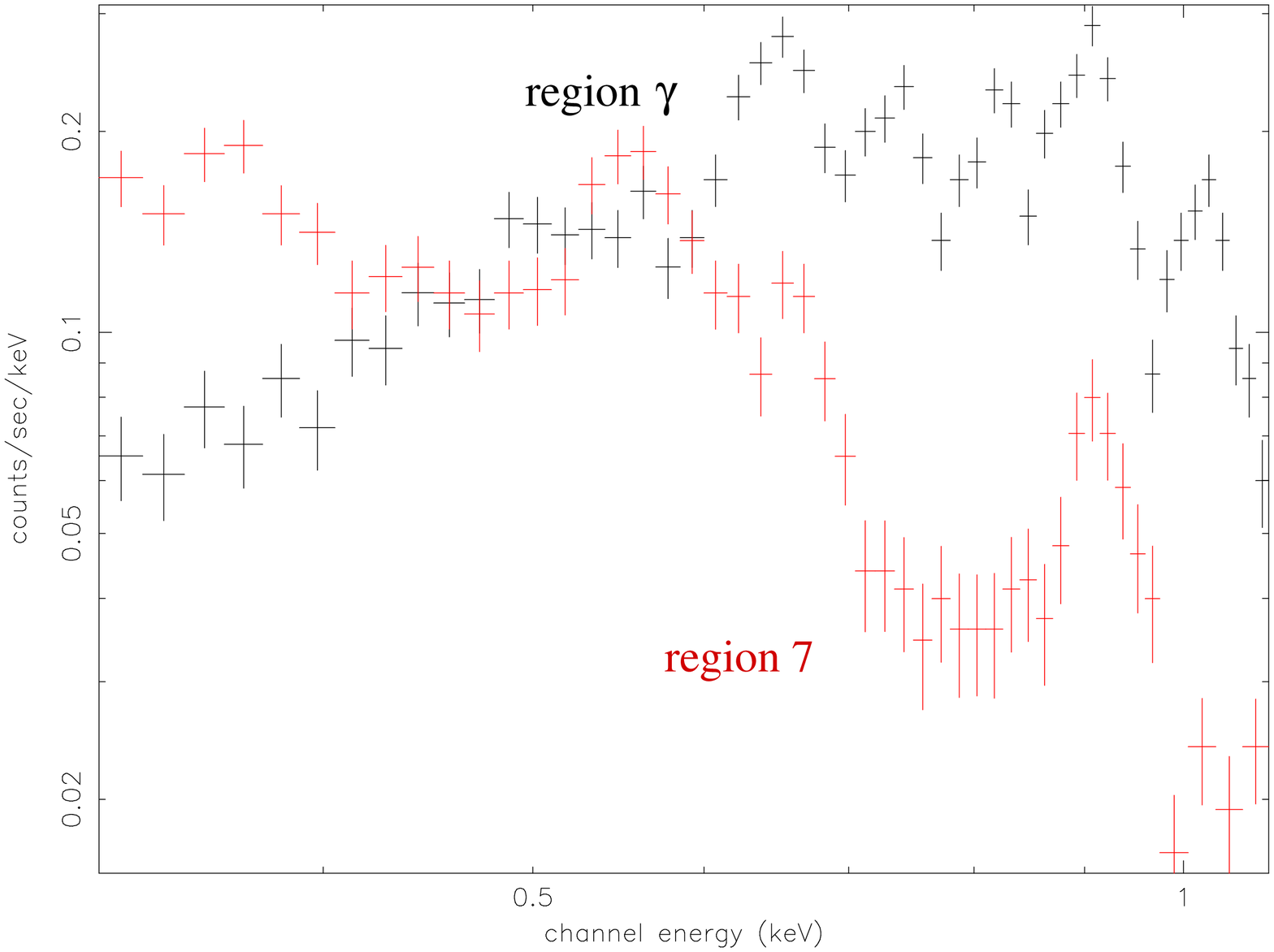,width=\columnwidth}	 
     }}
 \caption{\emph{Upper panel}: joint fit of the observed spectrum in region 2 (in red) and of the synthesized spectrum for setup Sph1 in region $\alpha$ (in black) with best-fit model and residuals. For the localization of the regions see Fig. \ref{fig:XMM} and Fig. \ref{fig:mappeXAB}. \emph{Lower panel}: Same as above, for region 7 and region $\gamma$.}
 \label{fig:spettriA}
 \end{figure}
 
\begin{figure}[htb!]
 \centerline{\hbox{
     \psfig{figure=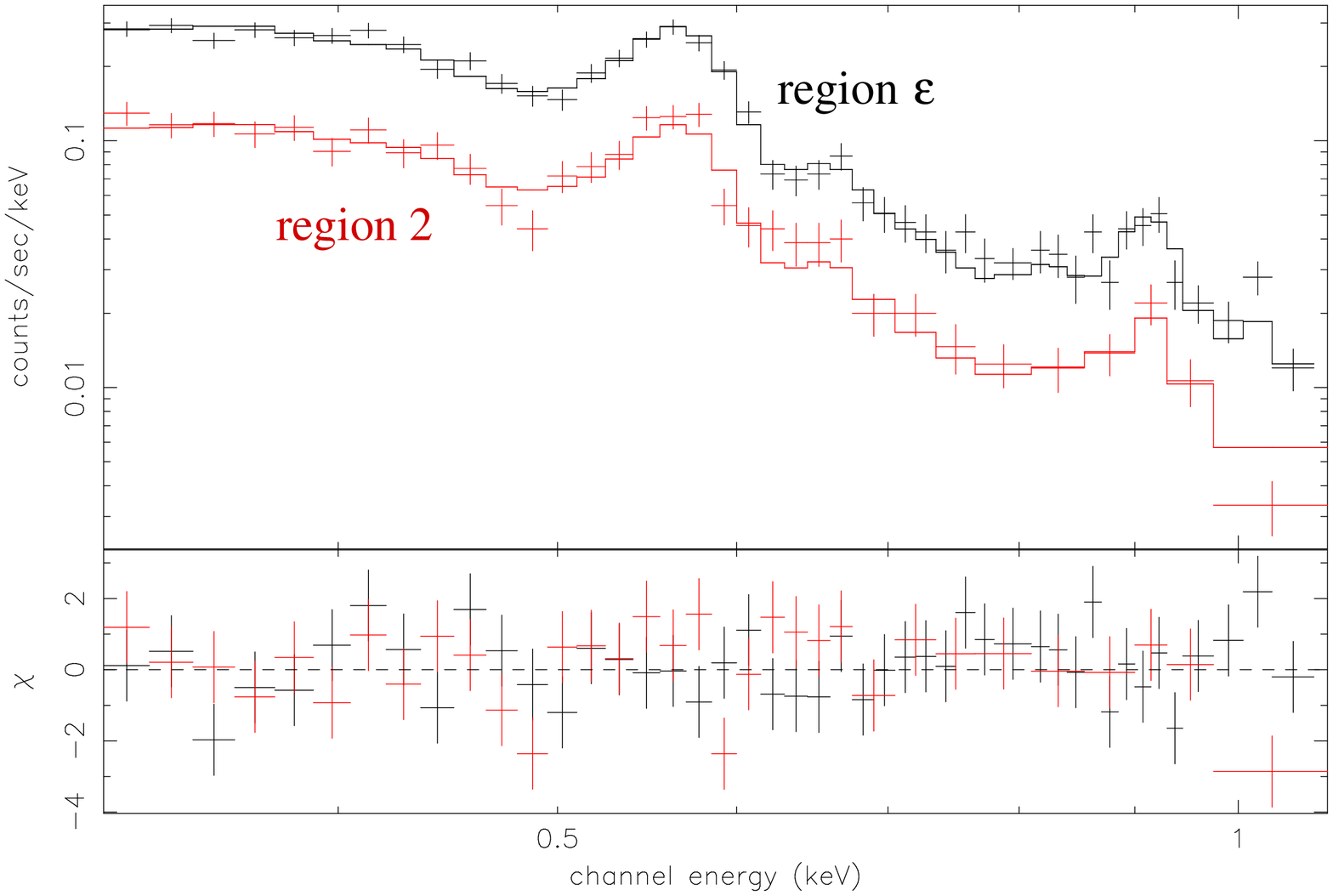,width=\columnwidth}	 
     }}
 \centerline{\hbox{
     \psfig{figure=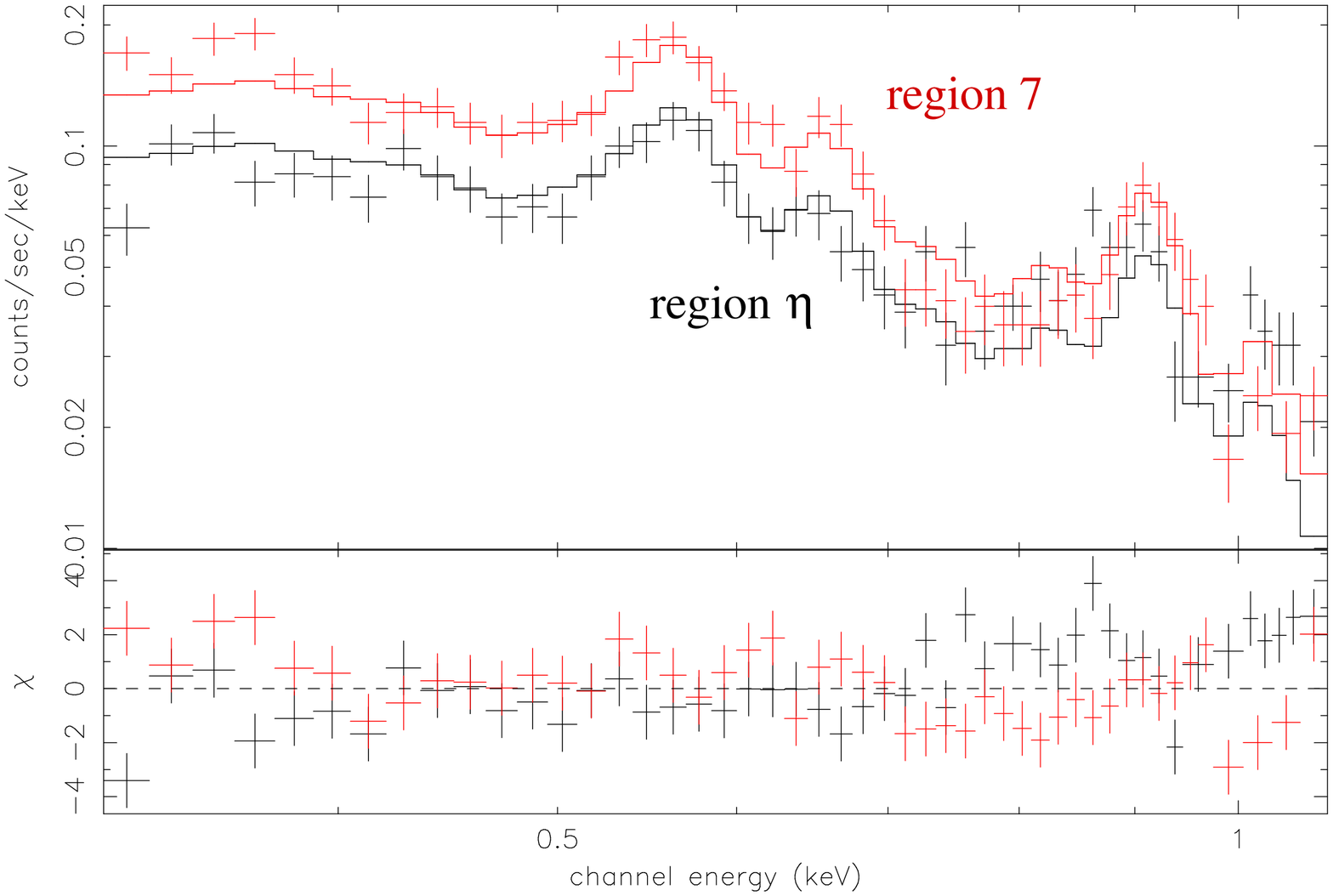,width=\columnwidth}	 
     }}
 \caption{\emph{Upper panel}: joint fit of the observed MOS spectrum in region 2 (in red) and of the synthesized MOS spectrum for setup Ell2 in region $\epsilon$ (in black). \emph{Lower panel}: Same as above, for region $\eta$ and region 7. The best fit models and residuals are also shown. For the localization of the regions see Fig. \ref{fig:XMM} and Fig. \ref{fig:mappeXAB}.}
 \label{fig:spettriB}
 \end{figure}
 
\begin{center}
\begin{table*}[htb!]
\begin{center}
\caption{Best-fit parameters of the spectral fitting performed simultaneously on the observed MOS spectra and on those synthesized from the hydrodynamic simulations for the spectral regions indicated in Fig. \ref{fig:XMM} and Fig. \ref{fig:mappeXAB}. The spectra are described with two thermal components (at temperature $T_{I}$ and $T_{II}$) in collisional ionization equilibrium, with an energy-independent multiplicative factor (first row) to take into account the differences in surface brightness between the synthesized and the observed spectra and the different areas of the spectral regions. All the reported errors are at 90\% confidence.}
\begin{tabular}{lccccc} 
\hline\hline
Parameter (setup Sph1$^{*}$)     &  Regions $\alpha-2$    &   Regions $\alpha-4$   &   Regions $\beta-4$    & Regions $\beta-7$     \\ \hline 
$factor$                      & $0.37\pm 0.02$         &    $0.45\pm0.02$       &  $0.30\pm0.02$         & $0.53^{+0.01}_{-0.03}$\\
$T_{I}$ ($10^{6}$ K)         & $0.99^{+0.02}_{-0.03}$ & $0.99^{+0.01}_{-0.04}$ & $1.44^{+0.06}_{-0.05}$ &    $1.39\pm{0.07}$    \\
$T_{II}$ ($10^{6}$ K)        & $3.4{+0.4}_{-0.3}$     &     $3.6\pm0.4$        &   $3.4{+0.2}_{-0.3}$   &    $3.2\pm0.2$        \\
$EM_{I}/EM_{II}$              &     $70\pm{20}$        &        $70\pm20$       &     $7.2\pm{1.3}$      & $4.8^{+0.8}_{-0.6}$   \\
$Fe/Fe_\odot$                 & $0.4^{+0.3}_{-0.1}$    &      $0.4\pm 0.1$      &      $0.3\pm 0.1$      & $0.6^{+0.2}_{-0.1}$   \\
Reduced $\chi^{2}$ (d. o. f.) &    $1.65$ $(67)$       &     $1.5$ $(69)$       &      $4.51$ $(78)$     &  $2.2$ $(91)$         \\ \hline \hline
Parameter  (setup Ell2)          & Regions $\delta-4$     &  Regions $\epsilon-2$  &    Regions $\zeta-6$   & Regions $\eta-7$      \\ \hline 
$factor$                      & $0.39^{+0.01}_{-0.02}$ &     $0.4 \pm 0.2$      &   $1.13^{+0.6}_{-0.5}$ & $1.43^{+0.1}_{-0.05}$  \\
$T_{I}$ ($10^{6}$ K)         & $1.11^{+0.05}_{-0.03}$ &    $1.04\pm{0.02}$     & $1.00^{+0.04}_{-0.06}$ &     $1.08\pm0.-5$     \\
$T_{II}$ ($10^{6}$ K)        &    $3.5 \pm 0.3$       &     $3.4\pm{0.2}$      &  $3.3^{+0.2}_{-0.4}$   &    $3.5 \pm 0.2$      \\
$EM_{I}/EM_{II}$              &    $40\pm 11$          &     $38_{-8}^{+10}$    &    $24_{-4}^{+5}$      &   $8\pm3$    \\
$Fe/Fe_\odot$                 &   $0.5\pm0.2$          & $0.4^{+0.2}_{-0.1}$    &   $0.5^{+0.2}_{-0.1}$  &     $0.4 \pm 0.1$    \\
Reduced $\chi^{2}$ (d. o. f.) &     $1.68$ $(65)$      &  $1.08$ $(72)$         &     $2.65$ $(80)$      & $2.07$ $(90)$ \\
\hline\hline 
\multicolumn{5}{l}{\footnotesize{* In region $\gamma$ the synthesized spectrum is completely different from the observed ones (reduced $\chi^{2}>10$).}} \\
\label{tab:spettriAB}
\end{tabular}
\end{center}
\end{table*}
\end{center}

To study the distribution of the photon energy across the field of view, we synthesized the MOS mean photon energy map (i.e., an image where each pixel holds the mean energy of the detected MOS photons) in the $0.3-2$ keV band. The map is shown in Fig. \ref{fig:avgEAB} (left panel), where we superimposed the surface brightness contour level at $75\%$ of the maximum in the same energy band. The brightest region does not correspond to the region with the minimum photon energy, at odd with what we observe in the FilD cloud (Fig. \ref{fig:avgEAB}, right panel).

In Paper I, we singled out spatial regions with homogeneous physical properties and we performed a spatially resolved spectral analysis on them. In order to compare the observed spectra with those synthesized from the hydrodynamic simulations, we have used the same procedure, by defining physically homogeneous regions. The regions are indicated in the upper right panel of Fig. \ref{fig:mappeXAB}, for setup Sph1. In each of these regions there are limited fluctuations of mean photon energy ($\la 1.7\%$ in region $\alpha$, $\la 6\%$ in region $\beta$, and $\la 2\%$ in region $\gamma$). Region $\alpha$ is in the bright northern part of the X-ray knot, where the emission associated to the transmitted shock dominates and the mean photon energy is low, region $\beta$ is in the brightest part of the cloud where both transmitted and reflected shocks contribute to the emission, and region $\gamma$ is located on the South, where we have high values of temperature and of mean photon energy. Part of the regions we selected for the analogous spatially resolved spectral analysis of the \emph{XMM-Newton} data are shown in Fig. \ref{fig:XMM}. The spectral fittings were performed simultaneously on the synthesized and the observed MOS spectra. In agreement with the findings of Paper I, we adopted a MEKAL model of an optically-thin plasma in CIE with two thermal components, we fixed $N_{H}=1\times 10^{20}$ cm$^{-2}$, and we left the model Fe abundance free and linked the Ne abundance to it, so as to have $(Ne/Ne_\odot)/(Fe/Fe_\odot)=4.4$. We added to the model an energy-independent multiplicative factor to take into account the differences in surface brightness between the synthesized and the observed spectra and the different areas of the spectral regions. Our results are summarized in Table \ref{tab:spettriAB}. Notice that, since the observed spectra are by themselves well described by this spectral model (as shown in Paper I), the $\chi^{2}$ values in Table \ref{tab:spettriAB} can be considered as an indication of the agreement between the observed and the synthesized spectra. As shown in the table and in the upper panel of Fig. \ref{fig:spettriA}, there is a good agreement between the spectrum synthesized in region $\alpha$ and those observed in the FilD regions with low mean photon energy (region 2 and region 4), but if we compare the spectra of region $\beta$ and region 4 (i. e. the ones with the highest synthesized and observed surface brightness), we have significant differences (see Table \ref{tab:spettriAB}). Instead, the spectrum of region $\beta$ has similar spectral features to the one observed in region 7, where the count rate is lower than in region 4, but the mean photon energy is higher. The spectrum extracted from region $\gamma$, not shown in the table, is completely different and significantly harder than all the observed spectra (see the lower panel of Fig. \ref{fig:spettriA}), even than those extracted form the RegNE cloud, which is the hardest X-ray emitting region in the EPIC field of view (see Paper I).

In \textbf{setup Sph2}, the post shock temperature of the cloud (i. e. the temperature behind the transmitted shock) is $\sim10^{6}$ K (as in setup Sph1), while the cloud density is slightly higher than in setup Sph1 ($n\la 6$ cm$^{-3}$). We synthesized the X-ray count-rate maps and focal plane spectra also for setup Sph2. The emission morphologies in the three energy bands ($0.3-0.5$ keV, $0.5-1$ keV, and $0.3-2$ keV) are very similar to those of setup Sph1. Moreover, the synthesized X-ray emission presents similar spectroscopic features to setup Sph1 (see Sect. \ref{A: spherical cloud}). However, the global X-ray luminosity of setup Sph2 is too high  (more than one order of magnitude) with respect to the observed one, therefore this setup will not be discussed in detail. 

\subsection{The ellipsoidal cloud}
\label{B: ellipsoidal cloud}

\textbf{Setup Ell1} differs from setup Sph1 for the shape of the cloud. However, for setup Ell1, we found that in the evolution of the shock-cloud interaction there are no conditions for optical emission of the plasma\footnote{We followed the evolution for $\sim 6000$ yr.}. In this case thermal instabilities do not develop and the plasma does not cool off to the characteristic temperatures of the optically emitting regions, while we observe a bright optical filament in the FilD cloud. Therefore, we will not discuss setup Ell1 further.

On the other hand, thermal instabilities leading to optically emitting regions develop in \textbf{setup Ell2}, that we discuss in detail in the following.
The lower panel of Fig. \ref{fig:nTAB} shows the 2D cross-sections through the $(r,~z)$ plane of temperature and density $\sim3600$ yr after the first impact of the shock front with the ellipsoidal cloud. As shown in Appendix A, this stage of the interaction is the one that best describes the observed data. The temperature and density cross-sections at different interaction times and the corresponding synthesized X-ray maps are available as online material.

At $t=3550$ yr, for setup Ell2, the plasma behind the transmitted shock has temperature $T\ga 10^{6}$ K and density $n\sim6-7$ cm$^{-3}$, the plasma behind the reflected bow shock has density $n\sim 0.4-0.5$ cm$^{-3}$ and it is about two times denser than the intercloud post-shock plasma. The pressure behind the transmitted shock front, $P^{t}$, the reflected shock front, $P^{r}$, and the main shock front, $P^{ms}$, are: $P^{t}\sim 2\times10^{-9}$ dyne cm$^{-2}$, $P^{r}\sim 7\times10^{-10}$ dyne cm$^{-2}$, $P^{ms}\sim 3\times10^{-10}$ dyne cm$^{-2}$. As shown in Fig. \ref{fig:nTAB}, the hot halo around the cloud due to its evaporation in the intercloud medium is also visible.

The lower panels of Fig. \ref{fig:mappeXAB} show the synthesized X-ray maps in the three energy bands, $0.3-0.5$ keV, $0.5-1$ keV, and $0.3-2$ keV, for setup Ell2. Also in this case the synthesized count rate per bin is consistent with the observed one (considering a $\sim30\%$ uncertainty for the distance of the Vela SNR), except for a small region in the northern edge of the synthesized X-ray knot. In the other parts of the knot the synthesized count rate per bin is closer to the observed one than setup Sph1. In setup Ell2 the hard X-ray emission from the plasma heated by the reflected shock does not contribute at large extent to the broadband emission. In this case, in fact, the X-ray emission in the $0.3-0.5$ keV band has almost the same morphology as those in the $0.5-1$ keV and $0.3-2$ keV bands, just like in our \emph{XMM-Newton} observation. Moreover, the overall morphology of the synthesized emission in all the energy bands is quite similar to the observed one.

Thermal instabilities occur also in this case, in the narrow sheets at the lateral boundary of the cloud $\sim3000$ yr after the shock impact. There the plasma rapidly cool off to temperatures of a few $10^{4}$ K, reaching density of a few tens cm$^{-3}$.  The lower panel of Fig. \ref{fig:X-HalphaS} shows the emission measure of the plasma with $T<10^{5}$ K and $n>3$ cm$^{-3}$ projected in the plane of the sky for setup Ell2. In this case the similarity with the observations is more evident: the impact of the Vela shock front with the ellipsoidal cloud produces an optical filament aligned with the shock velocity, with a morphology similar to the observed one.

Fig. \ref{fig:avgEAB} (central panel) shows the synthesized MOS mean photon energy map in the $0.3-2$ keV band for this setup. We superimposed the surface brightness contour level at $75\%$ of the maximum in the same energy band. In agreement with our observational findings (Fig. \ref{fig:avgEAB}, right panel), the brightest region corresponds to the region with the minimum photon energy.

As for the spatially resolved spectral analysis, we defined the four spatial regions with homogeneous physical properties shown in Fig. \ref{fig:mappeXAB} (lower right panel). In these regions the fluctuations of the mean photon energy are small ($\la 2\%$ in region $\delta$, $\la 7.5\%$ in region $\epsilon$ and $\zeta$, and $\la 5\%$ in region $\eta$). Region $\delta$ covers the brightest and softest northern part of the X-ray knot, region $\epsilon$ and $\zeta$ correspond to the central part of the cloud, and region $\eta$ is located at South, where we have higher values of temperature and of mean photon energy. We used the same spectral model described for setup Sph1 to analyze simultaneously the synthesized and the observed MOS spectra. Our results are summarized in Table \ref{tab:spettriAB}. As shown in the table and in Fig. \ref{fig:spettriB}, there is a good agreement between synthesized and observed spectra. If we compare the spectra of the brightest regions (i. e. region 4 and region $\delta$) we obtain a good agreement. Notice that region $\delta$ and region 4 are the ones with the minimum mean photon energy. We obtain a satisfying agreement also if we compare regions with higher mean photon energy.

\section{Discussion}
\label{Discussion}

We have pursued an accurate ``forward'' modeling of the interaction of the Vela SNR shock with an ISM cloud. The direct comparison of the observable synthesized from the simulations with the data is a powerful tool to constrain the model and to obtain insight in the physical conditions of the Vela region.

Our hydrodynamic model, set up from the analysis presented in Paper I, allows us to confirm that the bulk of the X-ray emission in the Vela FilD knot originates in the plasma behind the shock transmitted into an ISM cloud. We also confirm the important role of the thermal conduction between the shocked cloud and the intercloud medium and we demonstrate that thermal instabilities can explain the peculiar optical emission observed in the FilD region. We can ascertain the physical origin of the two spectral components we used in Paper I to model the observed X-ray spectra and we can also obtain information about the morphology of the ISM cloud.
Our hydrodynamic modeling shows, in fact, how the pre-shock morphology of the ISM cloud influences the dynamics of the shock-cloud interaction and the corresponding X-ray and optical emission. 

\begin{figure}[htb!]
 \centerline{\hbox{
     \psfig{figure=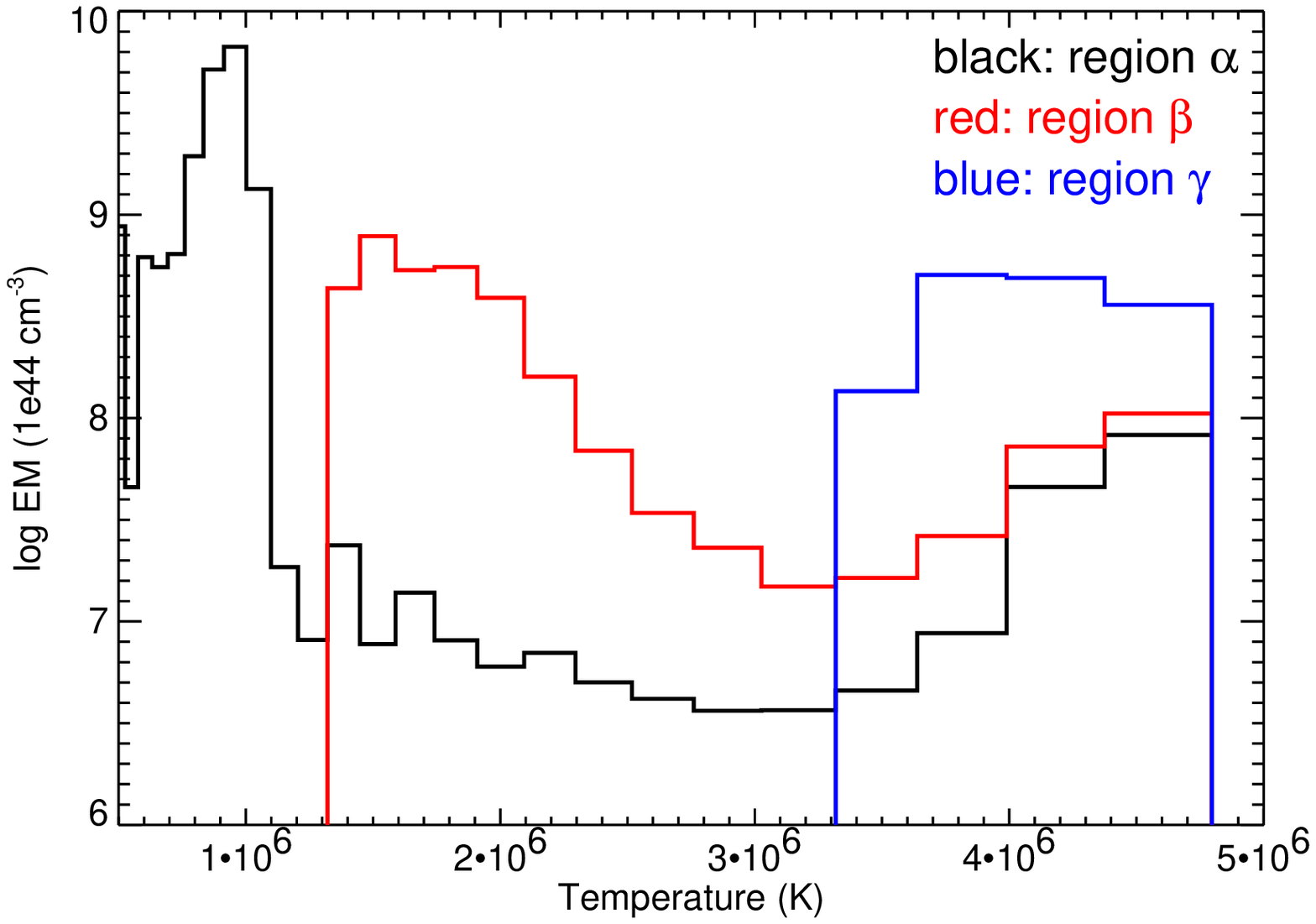,width=\columnwidth}	 
     }}
 \centerline{\hbox{
     \psfig{figure=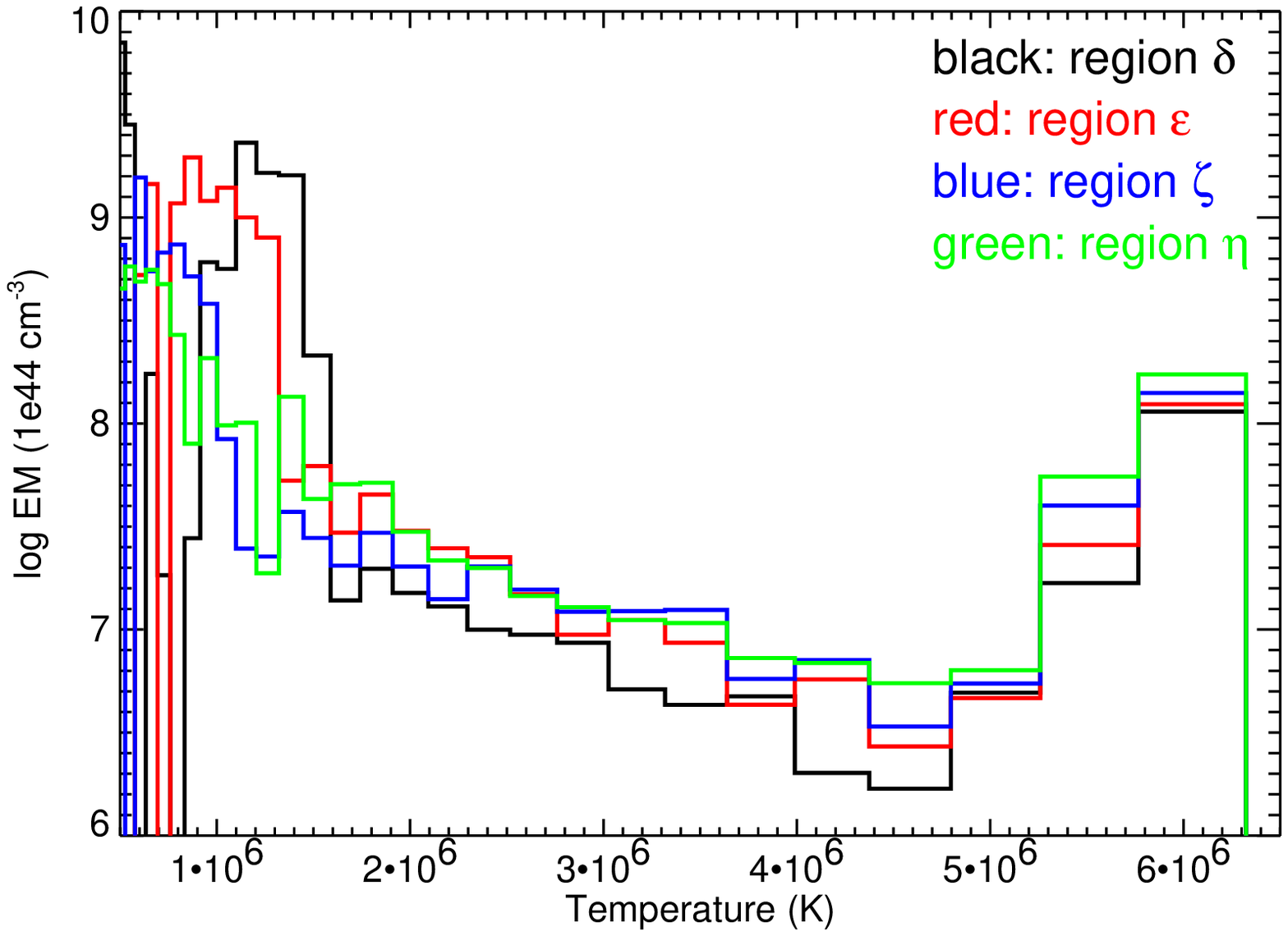,width=\columnwidth}	 
     }}
 \caption{Distribution of the emission measure vs. temperature in the spectral regions of Fig. \ref{fig:mappeXAB}, for the spherical cloud (setup Sph1, \emph{upper panel}) and the ellipsoidal cloud (setup Ell2, \emph{lower panel})}
 \label{fig:DEM}
 \end{figure}

Setup Sph1 and Ell2 essentially differ for the shape of the ISM cloud (spherical vs. ellipsoidal), with physical parameters adapted so as to have similar post shock temperature behind the transmitted shock.
Both in setup Sph1 and setup Ell2, the soft X-ray emission (below 0.5 keV) is associated to the cloud material heated by the transmitted shock front. In setup Sph1 a strong contribution to the X-ray emission above 0.5 keV comes from the intercloud material heated by the reflected shock and, in the end, the emission morphology and the spectral characteristics above 0.5 keV are not completely consistent with our \emph{XMM-Newton} observation of FilD. In particular, as shown in Fig. \ref{fig:mappeXAB}, above $0.5$ keV the bulk of the X-ray emission presents significant morphological differences from the softer emission (at odd with the observation) and the reflected shock emission is harder than the observed one (see lower panel of Fig. \ref{fig:spettriA}). Moreover, the optical filament is parallel to the plane of the shock front, at odd with the observation. Fig. \ref{fig:DEM} shows the distribution $EM$ vs. $T$ of the three regions of the upper right panel of Fig. \ref{fig:mappeXAB} for setup Sph1 (see also Tab. \ref{tab:spettriAB}). In region $\alpha$ the emission measure peaks at $\sim10^{6}$ K, that is the post-shock temperature of the cloud material, while the contribution of the intercloud material at $T\sim 4.6\times10^{6}$ K is about two orders of magnitude lower. In the central region $\beta$ the peak of the $EMD$ is at $\sim 1.5\times 10^{6}$ K and the contribution of the plasma with higher temperatures to the synthesized emission becomes considerable. In the southern region $\gamma$ all the emission originates in the hot intercloud plasma heated by the reflected shock front. We observe these features also for setup Sph2, which has the same shock temperature and cloud$/$intercloud density contrast as setup Ell2. Therefore, the models with a spherical cloud cannot explain all the spectral and morphological features we observe in FilD.

When the Vela blast wave shock impacts on an ellipsoidal cloud, the effects of the reflected shock to the X-ray emission are less considerable. In this case, in fact, the extension of the X-ray emitting knot in the $0.3-0.5$ keV band is similar to the one in the $0.3-2$ keV band (see Fig. \ref{fig:mappeXAB}), just like in the Vela FilD cloud.
Moreover, our model shows that thermal instabilities can explain the peculiar orientation of the FilD optical filament and its relationship with the X-ray emission. As for the X-ray surface brightness, both models yield higher values than the observed ones. This result may be partially due to distance effects\footnote{We remind that the distance $D$ of the Vela SNR is known with a $\sim30\%$ uncertainty and that the surface brightness scales as $D^{-4}$}. However, while in setup Sph1 the higher count rates are observed above $0.5$ keV, in setup Ell2 the brightest regions are those with the softest emission and the same happens in the FilD knot. 
The high surface brightness in setup Ell2 may be due to the too perfect alignment of the cloud axis and the main shock velocity, since the brightest knot corresponds to the region where the transmitted and the lateral shocks simultaneously interact. This extreme symmetry may also explain why in setup Ell2 we observe two vertical optical filaments, while in the FilD region the South-North filament is only one. In the case of the ellipsoidal cloud, all the synthesized spectra are remarkably in agreement with the observation. In all the spectral regions the soft component dominates and the surface brightness is related to its $EM$. These results are completely consistent with those obtained in Paper I. The lower panel of Fig. \ref{fig:DEM} shows the distribution $EM$ vs. $T$ of the four spectral regions selected for setup Ell2. In all the regions the contribution of the intercloud plasma heated by the reflected shock is constantly lower than that of the cloud material at lower temperature. 
It is worth noting that the spectral fittings performed on these four spectra have given results that are consistent with those obtained in Paper I. We conclude that setup Ell2 better describes the shock-cloud interaction processes in the Vela FilD region.

Our model results lead us to new physical insight for the interpretation of the \emph{XMM-Newton} data. In Paper I we found that the FilD spectra are described well by two thermal components at $T_{I}\sim 1\times 10^{6}$ K and $T_{II}\sim 3\times 10^{6}$ K. We associated these components with two different phases of the cloud: the cloud core and the cloud corona respectively. We can now better understand the physical origin of these two components. The cooler component clearly originates in the cloud material heated by the transmitted shock front. The cloud core is, in fact, associated to the narrow peak of the $EMD$ shown in the lower panel of Fig. \ref{fig:DEM}. Since the X-ray flux is always dominated by the cooler component, we conclude that in the Vela FilD knot the bulk of the X-ray emission originates behind the transmitted shock, which travels through an inhomogeneity of the ISM. The corona is also related to the cloud. As shown in Fig. \ref{fig:DEM}, we do not have a peak of the $EMD$ at $\sim 3\times 10^{6}$ K. The best-fit value $T_{II}$ instead reflects a wide temperature distribution, approximatively between $1.5\times10^{6}$ K and $5\times10^{6}$ K. This is the cloud plasma that ``evaporates'' because of the thermal conduction with the hotter intercloud medium (which corresponds to the secondary peak of the $DEM$) and forms a hot halo around the cloud. The X-ray emission from the ambient medium heated by the main shock is always negligible with respect to the core and corona emission. The X-ray emission from the ambient medium heated by the reflected shock front is barely visible below 0.5 keV, and, although visible in the $0.5-1$ keV band, is always significantly lower than the emission associated to the cloud.

Although our model is focussed on a particular region of the Vela SNR, and we have not performed a wide exploration of the parameter space, we can give a few indications for the interpretation of future shock-cloud observations: i) the contribution to the broad band emission of the intercloud medium behind the main shock front is negligible in all the explored setups. Therefore, the X-ray emission (even the high temperature component) is associated to the inhomogeneities of the ISM; ii) it is possible to ascertain how much the transmitted and reflected shocks contribute to the X-ray emission, even if the statistics is not enough for a spatially resolved spectral analysis, through the comparison between the count rate image and the mean photon energy map in a wide energy band: if the mean photon energy map (Fig. \ref{fig:avgEAB}) has a minimum in the regions with the highest surface brightness, the emission behind the transmitted shock front dominates, otherwise the contribution of the plasma behind the reflected shock is significant; iii) a typical approach for the determination of the plasma density of the X-ray emitting plasma from the spectral analysis is based on the assumption of pressure equilibrium between the different spectral components (this allows to derive, for each component, the filling factor and the density from $EM$ and $T$). Our hydrodynamic model shows that this assumption may not be fully hold. However, the pressure variations are within a factor of 5, for the spherical cloud, and a factor of 10, for the elliptical cloud. 

We also obtained indications about the role of the different physical processes in the dynamics of the shock-cloud interaction. In particular, for example, the radiative losses determine the formation of thermal instabilities and the consequent optical emission, while thermal conduction between the cloud and the intercloud medium induces the evaporation of the cloud and the formation of the X-ray emitting ``corona''.

The efficiency of thermal conduction may be significantly reduced by the presence of the magnetic field (which is not taken into account in our model). If we assume an organized ambient magnetic field, the thermal conduction is anisotropic, because the conductive coefficient in the direction perpendicular to the field lines is several orders of magnitude lower than that parallel to the field lines, which coincides with the Spitzer's coefficient $\kappa_{spi}$.  In the presence of a pre-shock uniform magnetic field, the ambient magnetic field is expected to envelope the shocked cloud during the shock-cloud interaction (\citealt{fag05}, where the effects of thermal conduction are not taken into account) and this prevents the evaporation of the cloud under the effect of thermal conduction and promotes the formation of thermal instabilities.

However, \citet{nm01} have shown that, in the presence of a multiscale turbulent magnetic field, the efficiency of turbulent thermal conduction approaches the Spitzer's limit, the diffusion constant being $\sim \kappa_{spi}/5$. Since the conductive flux is proportional to $T^{7/2}$, this reduction of the efficiency of thermal conduction can be counterbalanced by increasing the shock temperature by a factor of $5^{2/7}\sim 1.58$, and/or by slightly reducing the cloud/intercloud density contrast. Since our model well reproduces the observed data, we conclude that there is no need to include an organized ambient magnetic field, but our results substantially hold in the presence of a turbulent magnetic field.

\section{Summary and conclusions}
\label{Summary and conclusions}
Our analysis is focussed on the physical description of a bright isolated X-ray knot in the northern rim of the Vela SNR (Vela FilD), whose X-ray emission has been studied and discussed in Paper I.
We modelled the hydrodynamics of the interaction between a SNR blast wave shock and an isolated cloud, taking into account thermal conduction and radiative cooling effects. We explored four different model setups, choosing the values of the physical parameters on the basis of our preliminary analysis of the \emph{XMM-Newton} X-ray data. Since we synthesized the X-ray emission filtered through the \emph{XMM-Newton} EPIC-MOS instrumental response from the model, we can compare the model results directly to the data and derive detailed and quantitative information and deep insight. Our study has shown that:
\begin{itemize}
\item The dynamics of the shock-cloud interaction and the corresponding X-ray and optical emission are significantly influenced by the pre-shock morphology of the ISM cloud. In particular, if the Vela shock front impacts on a spherical cloud, both transmitted and reflected shocks contribute to the X-ray emission, while in the case of an ellipsoidal cloud, the reflected shock contribution is negligible.
\item An ellipsoidal cloud 30 times denser than the intercloud medium, with the semimajor axis parallel to the shock velocity describes well the \emph{XMM-Newton} observation of the Vela FilD knot in terms of X-ray emission spectral properties and morphology, and in terms of spatial relationship between optical and X-ray emission.
\item The peculiar orientation of the FilD optical filament (which is difficult to explain according to the common scenario which associates optical emission in SNRs with slow shocks travelling through dense clouds) is naturally explained by this setup as a result of thermal instabilities.
\item Our model allows us to understand the physical origin of the two thermal components we used in Paper I to describe the FilD observed spectra. The cooler component originates in the cloud material heated by the transmitted shock front, while the hotter component is the result of thermal conduction between the cloud and the hotter shocked intercloud medium. The heating of the cloud produces its evaporation and the formation of a hot diffuse halo. 
\item We highlighted spectral and morphological characteristics of the emission from transmitted and reflected shocks, thus obtaining a key for the interpretation of future shock-cloud observations.
\end{itemize}

Future studies should also address other open issues, like the presence of a single optical filament observed in FilD and the effects of the pre-shock symmetry of the cloud on the morphology of the optical emission and the on the surface brightness in X-rays. Moreover, we also aim at describing the shock-cloud interaction in RegNE, which is an other bright X-ray knot, hotter than FilD, discussed in Paper I.

\begin{acknowledgements}
We thank the referee, A. Bykov, for his comments and
suggestions. The software used in this work was in part developed by the DOE-supported ASC / Alliance Center for Astrophysical Thermonuclear Flashes at the University of Chicago. The simulations were performed on the IBM/Sp4 machine at CINECA (Bologna, Italy) in the framework of the INAF / CINECA Agreement, and on the Compaq cluster and the Linux cluster at the SCAN facility of the INAF$-$Osservatorio Astronomico di Palermo. This work was partially supported by the Ministero dell'Istruzione, dell'Universit\`{a} e della Ricerca and by Agenzia Spaziale Italiana.
\end{acknowledgements}

\bibliographystyle{aa}
\bibliography{references}

\appendix 

\section{Determination of the evolutionary stage of the interaction}

The evolution of the shock-cloud interaction has been followed for $\sim 6000$ yr, for all the model setups presented in this paper. We have developed a procedure for the selection of the evolutionary stage of the interaction which best reproduce the observed data. This procedure is based on the definition of three quantities: 
\begin{enumerate}
\item The extension, $A$, of the X-ray emitting knot, defined as the area covered by the region with an X-ray surface brightness three times larger than the minimum (in the $0.3-2$ keV energy band).
\item The mean photon energy, $MPE$, of the X-ray emitting knot, defined as the mean photon energy in the X-ray emitting region with an X-ray surface brightness three times larger than the minimum (in the $0.3-2$ keV energy band).
\item The count-rate, $R$, in the soft region, defined as the average count-rate in the region with a mean photon energy between the minimum $E_{min}$ and $1.2\times E_{min}$.
\end{enumerate}

\begin{figure}[htb!]
 \centerline{\hbox{
     \psfig{figure=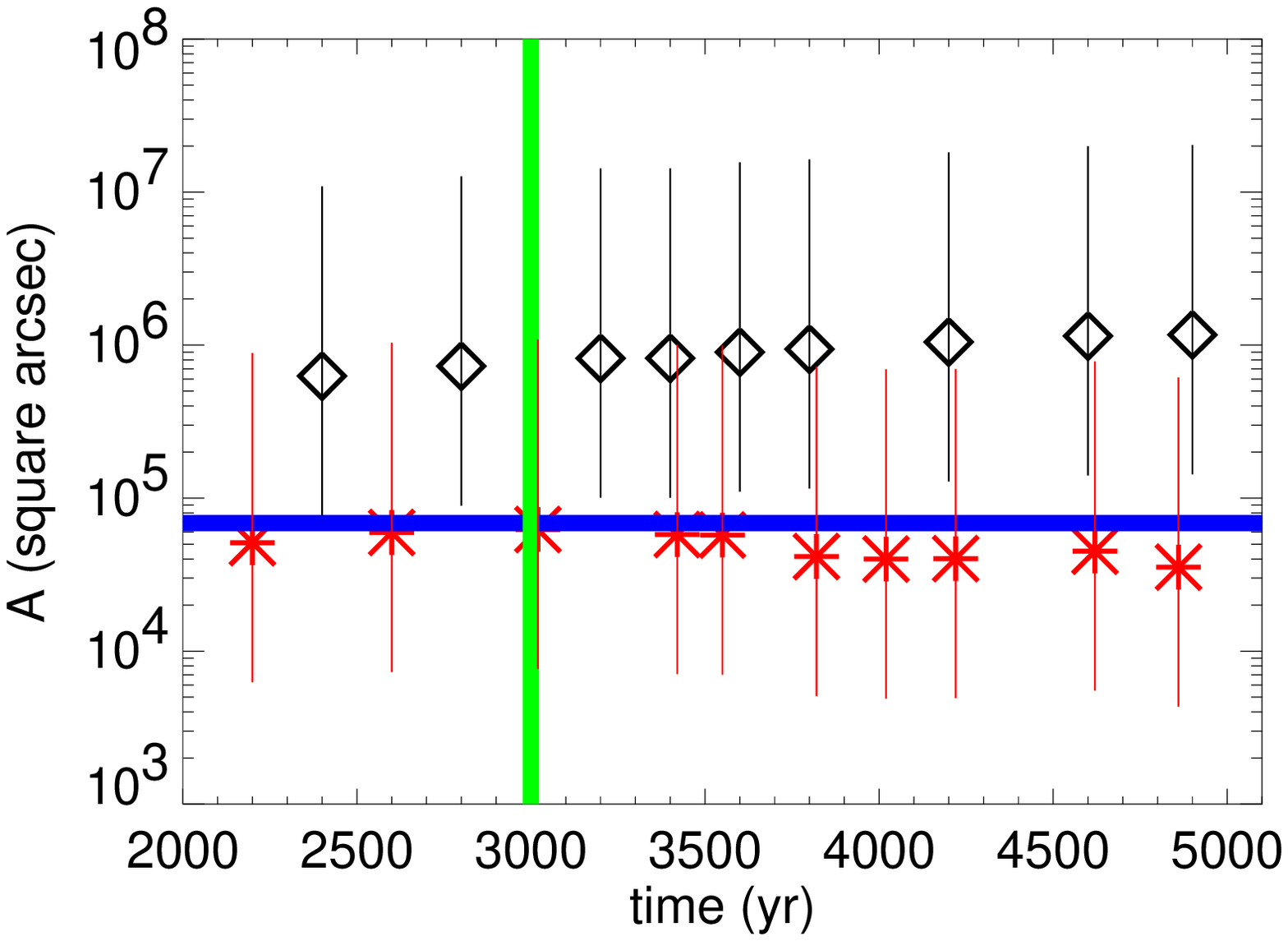,width=\columnwidth}	 
     }}
 \centerline{\hbox{
     \psfig{figure=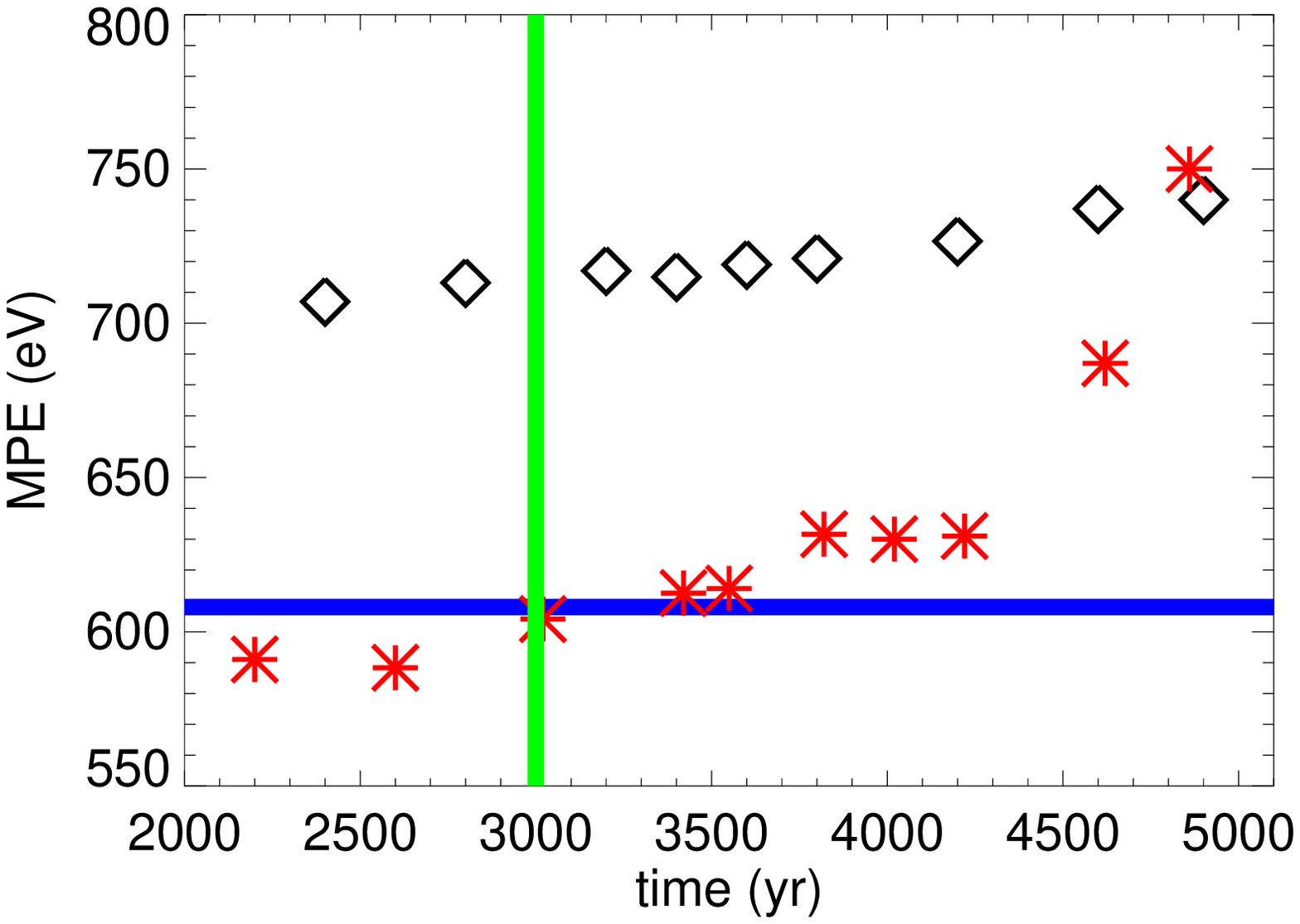,width=\columnwidth}	 
     }}
 \centerline{\hbox{
     \psfig{figure=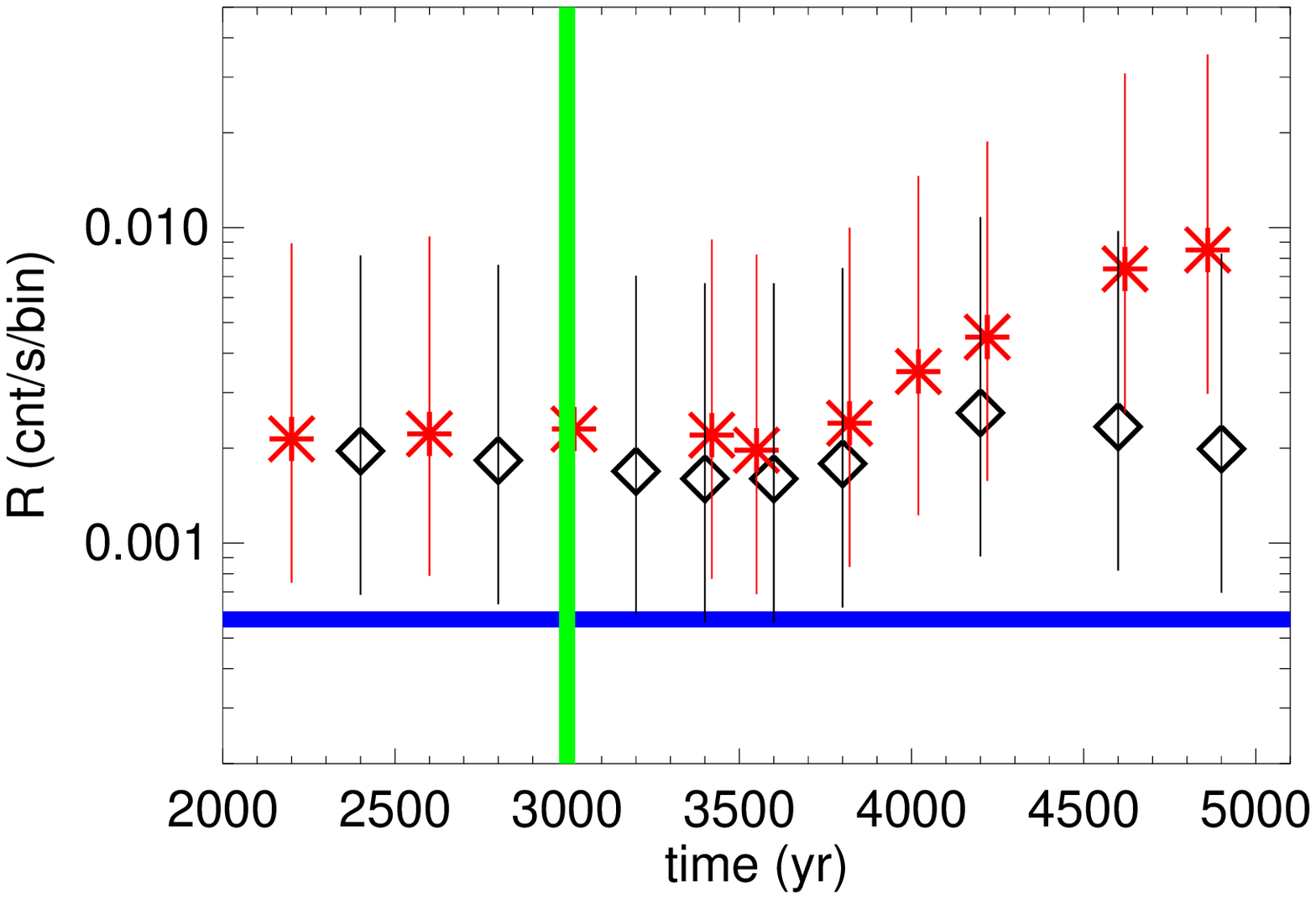,width=\columnwidth}	 
     }}
 \caption{Temporal evolution of the quantities $A$, \emph{upper panel}, $MPE$, \emph{central panel}, and $R$ \emph{lower panel} (defined in Appendix A), for setup Sph1 (black diamond) and setup Ell2 (red cross). The error bars are associated with the $\sim30\%$ uncertainty in the distance of the Vela SNR. The blue lines correspond to the observed values and the green lines indicate the epoch of the formation of thermal instabilities.}
 \label{fig:timesel}
 \end{figure}

Figure \ref{fig:timesel} shows the temporal evolution ($t=0$ corresponds to the impact of the shock front with the cloud) of $A$ (upper panel), $MPE$ (central panel), and $R$ (lower panel), for setup Sph1 (black diamond) and setup Ell2 (red cross). The corresponding observed values are indicated by the horizontal blue lines, while the vertical green lines indicate the formation of thermal instabilities: no optical emission is therefore present on the left of the green line. 

The graphs in Fig. \ref{fig:timesel} indicate that, for both setups, we can identify a temporal range, around $3500\pm500$ yr where the synthesized values globally best approach the observed ones. Below $t=3000$ yr we do not have optical emission, while above 4000 yr the synthesized values have largest deviations from the observed ones. For all the three quantities, the smallest deviations are at $\sim 3500$ yr, therefore in this paper we report the results at $t=3400$ yr, for setup Sph1, and $t=3550$, for setup Ell2. 

Figure \ref{fig:timesel} also indicates that, for $t\sim3500$ yr, setup Ell2 well reproduces the observed values of $A$ and $MPE$, while in setup Sph1 we have a large area of the X-ray emitting knot and a mean photon energy in the bright regions that is larger than the observed one. This is in agreement with the results discussed in Sect. \ref{Results}, where we show that in setup Sph1 the bright regions do not correspond to the minima in the mean photon energy map. As for setup Ell2, the high values of the count-rate in the soft region are due to the brightest part in the northern edge of the X-ray knot (see Fig. \ref{fig:mappeXAB}), where we have the minimum in the mean photon energy map (Fig. \ref{fig:avgEAB}, right panel).
\end{document}